\documentclass[twocolumn,prl,aps,floatfix,superscriptaddress]{revtex4-1}
\usepackage{bm,color,amsmath,amssymb,mathrsfs,latexsym,graphicx,psfrag,float}
\usepackage[caption=false]{subfig}
\usepackage{times}

\newcommand{\be}{\begin{equation}}
\newcommand{\ee}{\end{equation}}
\newcommand{\bea}{\begin{eqnarray}}
\newcommand{\eea}{\end{eqnarray}}

\renewcommand{\phi}{\varphi}
\renewcommand{\epsilon}{\varepsilon}
\renewcommand{\vec}[1]{{\bf #1}}

\newcommand{\CMG}{{Co$_2$MnGa}}

\begin{document}
\fontsize{.34cm}{.35cm}\selectfont

\title{Enhanced higher harmonic generation from nodal topology}

\affiliation{School of Physics, Sun Yat-sen University, Guangzhou 510275, China.}
\affiliation{Instiute of High Performance Computing, A*STAR, Singapore, 138632}
\affiliation{Department of Physics, National University of Singapore, Singapore, 117542.}
\affiliation{NUS Graduate School for Integrative Sciences and Engineering, Singapore 117456, Republic of Singapore.}
\affiliation{University of Cambridge, Cambridge, United Kingdom.}
\affiliation{Wuhan National High Magnetic Field Center and School of Physics, Huazhong University of Science and Technology, Wuhan 430074, China.}

\author{Ching Hua Lee}
\email{calvin-lee@ihpc.a-star.edu.sg}
\affiliation{Instiute of High Performance Computing, A*STAR, Singapore, 138632}
\affiliation{Department of Physics, National University of Singapore, Singapore, 117542.}

\author{Han Hoe Yap}
\affiliation{Department of Physics, National University of Singapore, Singapore, 117542.}
\affiliation{NUS Graduate School for Integrative Sciences and Engineering, Singapore 117456, Republic of Singapore.}

\author{Tommy Tai}
\affiliation{University of Cambridge, Cambridge, United Kingdom.}

\author{Gang Xu}
\affiliation{Wuhan National High Magnetic Field Center and School of Physics, Huazhong University of Science and Technology, Wuhan 430074, China.}

\author{Xiao Zhang}
\email{zhangxiao@mail.sysu.edu.cn}
\affiliation{School of Physics, Sun Yat-sen University, Guangzhou 510275, China.}

\author{Jiangbin Gong}
\email{phygj@nus.edu.sg}
\affiliation{Department of Physics, National University of Singapore, Singapore, 117542.}

\date{\today}

\begin{abstract}
\end{abstract}


\maketitle
\textbf{Among topological materials, nodal loop semimetals (NLSMs) are arguably the most topologically sophisticated, with their valence and conduction bands intersecting along arbitrarily intertwined nodes. But unlike the well-known topological band insulators with quantized edge conductivities, nodal loop materials possess topologically nontrivial Fermi surfaces, not bands. Hence an important question arises: Are there also directly measurable or even technologically useful physical properties characterizing nontrivial nodal loop topology? In this work, we provide an affirmative answer by showing, for the first time, that nodal linkages \emph{protect} the higher harmonic generation (HHG) of electromagnetic signals. Specifically, nodal linkages enforce non-monotonicity in the intra-band semi-classical response of nodal materials, which will be robust against perturbations preserving the nodal topology. These nonlinearities distort incident radiation and produce higher frequency peaks in the teraHertz (THz) regime, as we quantitatively demonstrate for a few known nodal materials. Since THz sources are not yet ubiquitous, our new mechanism for HHG will greatly aid applications like material characterization and non-ionizing imaging of object interiors~\cite{terahertzApplication,terahertzApplication2}.
}

About a decade ago, topological materials were born when signatures of quantum spin Hall (QSH) states were detected in HgTe quantum wells~\cite{konig2007quantum}. Since then, an intensive search has returned various alternative topological material candidates~\cite{jacutingaite,QAH_LaCl} with promise in applications like dissipationless wire interconnects~\cite{mellnik2014spin,zhang2015electrical,zhang2016autobahn,zhang2016topologicalIC}. More recently, the scope of topological materials has expanded to also include gapless 3D nodal systems like Weyl and nodal line semimetals, where the \emph{Fermi surface} itself is topologically nontrivial~\cite{nodalSurface,NLSM_FuLiang,chen2017topological,PhysRevB.96.041103,graphdiyne}. In particular, nodal lines have been experimentally observed in PbTaSe$_2$~\cite{bian2016topological}, BiTeI~\cite{crepaldi2012giant}, ZrSiTe, ZrSiSe~\cite{hu2016evidence}, the centrosymmetric superconductor SnTaS$_2$~\cite{SnTaS2} and, in the form of nodal chains, TiB$_2$~\cite{yi2018observation}. 

Compared to conventional topological insulators which fall into $\mathbb{Z}_2$ or $\mathbb{Z}$ topological classes~\cite{qi2008,classification}, NLSMs possess far richer topology, with their nodal loops knotted or linked in unlimited topologically distinct ways~\cite{AlexanderPolynomial,vortexKnot,lee2018tidal}.  A NLSM consists of valence and conduction bands intersecting along so-called ``nodal lines'' in 3D momentum space. When these band intersections occur at approximately constant energy, a small chemical potential will result in Fermi regions in the shape thin ``tubes'' along the nodal lines, even if a small gap prevents perfect band degeneracy. Depending on the crystal symmetry, these Fermi tubes can close to form nodal loops (NLs) that touch to form nodal chains, or link/entangle among themselves to form nodal links/knots (Fig.~\ref{fig:materials}), as proposed in $ABC$-stacked graphdiyne~\cite{graphdiyne} and Co$_2$MnGa~\cite{CMG_ARPES}. Even with only one nodal loop, the possible topologies are already uncountably infinite: the loop can cross itself from above or below any number of times, with each permutation leading to a multitude of other knotted configurations. With three or more loops, the nodal structure can even be non-trivially linked despite having no pairwise linkage, as in a Borromean-ring configuration characterized by the Milnor topological invariant~\cite{trang1976invariance,murasugi2007knot,staalhammar2019hyperbolic}. Such is the richness of nodal topologies that no single topological invariant can unambiguously distinguish between all topologically distinct configurations.

While quantized Hall conductivity is the hallmark of nontrivial band topology in topological insulators, it is so far unclear what distinct experimental signature corresponds to nontrivial nodal loop topology~\footnote{More precisely, the topological class of the Fermi region configuration around the nodal loops.}. 
The existence of a nodal loop and its surface states is known to give rise to weak anti-localization~\cite{chen2019weak} and spin-resolved transport~\cite{chen2018proposal} signatures respectively, but these properties cannot resolve the topology of nodal linkages. Progress has been made through Berry phase interference measurements around nodal structures, which yield homotopy data that can be used to map out the nodal topology~\cite{li2016bloch,flaschner2016experimental}. However, such experiments involve intricately specified paths whose very design require a priori knowledge of the nodal structure. Furthermore, they are impractical except in nodal cold atom systems, for which probing techniques like Bloch state tomography and center-of-mass response measurements are applicable and mature~\cite{BlochStateTomography}. For characterizing actual NLSM materials, the only experimental recourse so far had been ARPES measurements~\cite{CMG_ARPES,bian2016topological}, although they are arguably indirect approaches involving extensive data reconstruction. Only with a more definitive experimental signature can the mathematical appeal of nodal topology be elevated to phenomenological significance.


In this work, we show that this sought-after topological signature can be found in the protection of HHG in NLSMs. Specifically, we demonstrate that nodal loop linkages impose lower bounds on the nonlinearity of their intraband optical response, hence ensuring robust HHG as long as the nodal topology remains undisturbed. While nodal points and lines 
are already known to exhibit interesting HHG and magneto-optical responses due to their peculiar density of states~\cite{Morimoto,Zyuzin}, no existing works have shown how these properties can possibly be protected by topology. Indeed, existing studies have typically been  perturbative, unable to access the non-perturbative regime where the incident field impulse is large enough to probe the nodal linkage. Through a semi-classical Boltzmann approach, we shall explore the effects of arbitrarily strong fields, and demonstrate generic significant enhancement of HHG by nodal topology. For a few experimentally relevant (slightly gapped) NLSMs, namely Ti$_3$Al, YH$_3$ and Co$_2$MnGa, we shall also provide quantitative estimates of this enhancement contribution.

\noindent\textbf{Nonlinear semi-classical response --} We first describe the semi-classical Boltzmann approach, from which the response curves of a 3D electronic material can be determined from the shape of its Fermi region. Similar semi-classical approaches have been highly successful in explaining phenomena like Hall effects and quantum oscillations in diverse settings~\cite{xiao2010berry}, as well as Bloch oscillations and Berry curvature effects in the context of HHG~\cite{schubert2014sub,liu2017high,SuppMat}. Subject to an external $\bold E$ field, the response current is given by~\footnote{normalization implicit in integration measure} 
\begin{equation}
\bold J = \int d^3\bold k\,g(\epsilon(\bf k)) \langle \bf k|J|\bf k\rangle,
\end{equation}
where $g(\epsilon(\bf k))$ is the non-equilibrium occupation function that depends explicitly on the energy dispersion $\epsilon(\bf k)$, and implicitly on $\bold E$ through $g$. The semi-classical assumption is that the collective effect of a multitude of possible many-body processes can be simply encapsulated by the behavior of $g(\epsilon(\bf k))$ and $\langle \bf k|J|\bf k\rangle$, which holds as long as the electrons are well-described by a Fermi liquid~\cite{kubo1973boltzmann}. Under the constant relaxation time approximation, the electronic occupation function $g$ satisfies the Boltzmann's equation 
\begin{equation}
\frac{dg}{dt}=\frac{\partial g}{\partial t}+\bold{\dot k}\cdot \frac{\partial g}{\partial \bf k}=\frac{f-g}{\tau},
\label{Boltzmann}
\end{equation}
such that it 'relaxes' to the local equilibrium (Fermi-Dirac) distribution $f=F(\epsilon(\bold k))=(1+e^{\beta \epsilon(\bf k)})^{-1}$ at a rate inversely proportional to the relaxation time $\tau$, which can be computed from a microscopic model for the scattering processes~\cite{ho2018theoretical}. Physically, $\frac{d}{dt}=\frac{\partial}{\partial t}+\bold{\dot k}\cdot \frac{\partial }{\partial \bf k}$ is the phase space convective derivative, which when acted on $g$ gives the correction $(f-g)/\tau$ to the continuity equation due to scattering. We have omitted spatial dependencies $\frac{\partial g}{\partial \bf r}$ since we are not considering thermal or chemical gradients. 

The effect of an external field $\bold E$ enter the semi-classical equations of motions 
for electronic wavepackets, which take a manifestly symmetric form when expressed in units where electronic mass, charge and $\hbar$ are all set to unity:
\begin{subequations}
\begin{align}
\bold{\dot r}&= \bold v+\bold{\dot k}\times \bold \Omega
\label{EOMr}\\
\bold{\dot k}&= \bold E + \bold{\dot r}\times \bold B.
\label{EOMp}
\end{align}
\end{subequations}
Here, $(\bold r,\bold k)$ are the center-of-mass phase space coordinates of an electron wavepacket. In this work, we shall not consider the effects of a magnetic field, and set $\bold B=\bold 0$. The wavepacket group velocity $\bold v =\boldsymbol{\nabla_k}\epsilon(\mathbf{k})$ and Berry curvature $\bold \Omega=\bold \Omega(\bold k)$ it feels are both explicit functions of the wavepacket momentum $\bold k$. 
Due to translation invariance implied by Eq.~\ref{Boltzmann}, $g$ is only affected by the Lorentz force $\bold{\dot k}$, and not $\bold {\dot r}$.
\begin{figure*}
\begin{minipage}{\linewidth}
\includegraphics[width=\linewidth]{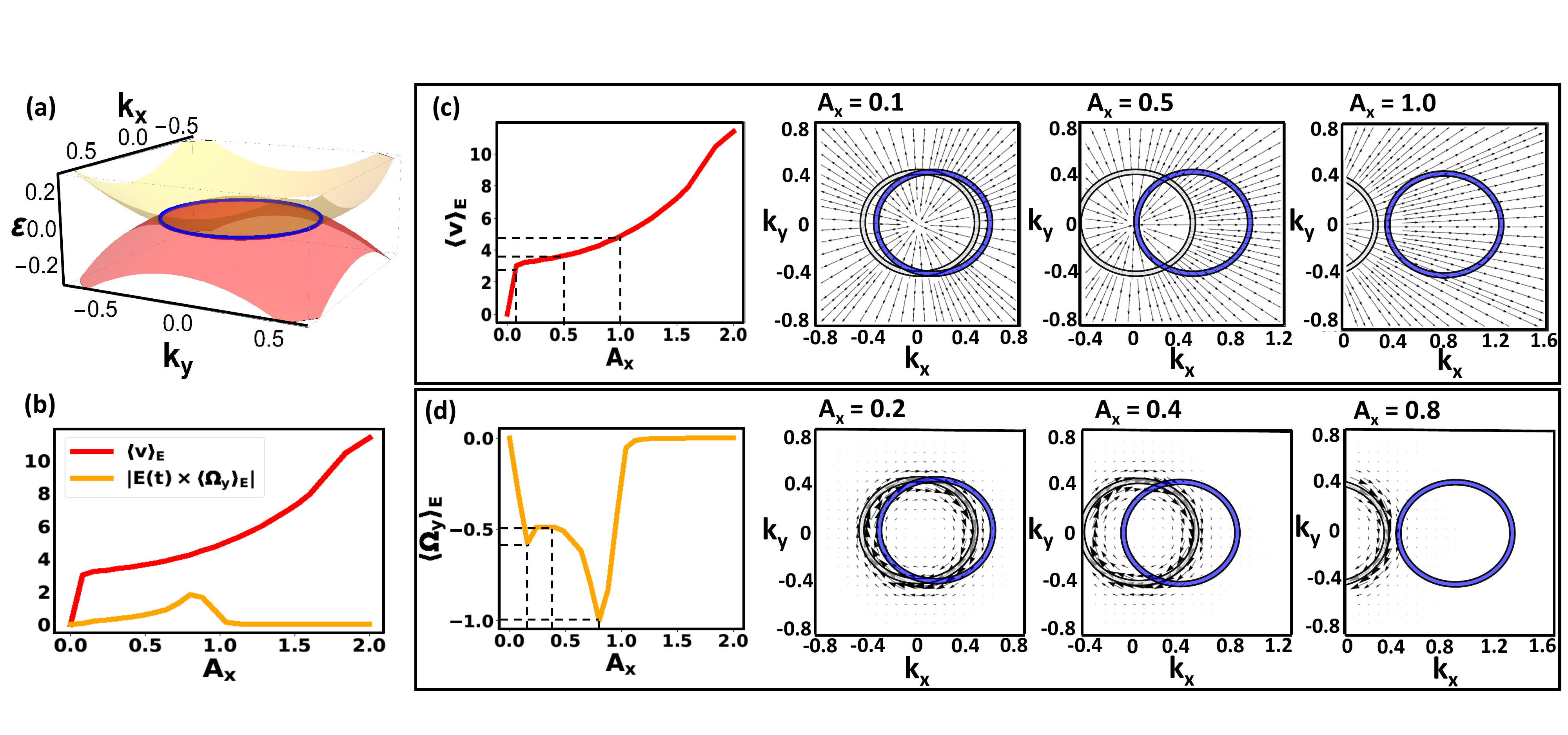}
\end{minipage}
\caption{Unprotected nonlinear response of a topologically trivial single NL. (a) Illustrative energy dispersion in the $k_z=0$ slice defined by Eq.~\ref{H_2} with $\mu=0.03$, $r=2.6$ and $T=10$K. (c) $\langle v_x\rangle_{\bold E}$ contribution, in arbitrary units, to the longitudinal response from interferences in velocity vector field. The partial cancellation of $\bold v =\boldsymbol{\nabla_k}\epsilon(\mathbf{k})$ at electromagnetic vector potential $A_x\sim E_x/\Omega$ comparable to the NL radius (center) suppresses the increase of $J_x(A_x)$, but not sufficiently strong to create a response kink (red curve). (d) Transverse $\langle \Omega_y\rangle_{\bold E}$ contribution to the response from interferences in Berry flux field, shown for $r=2.9$. Since the Berry curvature is strongly concentrated around the NL, the response peaks only when the Fermi tube overlaps considerably with the NL. The smaller peak at $A_x\approx 0.1$ is, however, partially suppressed by opposite contributions from both sides of the NL. (b) Resultant current response from both contributions, which is still monotonic despite being nonlinear. The Berry curvature contribution is typically subdominant for illustrative parameters drawn from realistic nodal ring materials. }
\label{fig:ring}
\end{figure*}

To derive the explicit response dependence on the nodal structure, we shall focus on the ballistic regime where $\tau$ dominates all other timescales, such that closed form solutions exist for $g$ and hence the response current $\bold J$. In particular, this requires that $\Omega\tau\gg 1$, where $\bold E(t)\sim \bold E(0)e^{i\Omega t}$, a reasonable constraint in the THz regime ($\Omega\sim 20-100$ THz) where $\Omega\tau$ can be estimated to lie between $10$ to $50$ for known nodal materials~\cite{SuppMat}. In this ballistic regime, scattering processes cannot catch up with the much shorter oscillation timescales~\cite{liu2017high,ndabashimiye2016solid}, and we do not expect $g$ to be close to the local equilibrium $f$ at all; instead, $g$ obeys the very intuitively attractive solution
\begin{equation}
g(\bold k,t)=F\left(\epsilon\left(\bold k-\int^t_{-\infty}\bold E(t')dt' \right)\right)=F\left(\epsilon\left(\bold k-\bold A(t) \right)\right),
\label{ballistic}
\end{equation}
which can be easily verified viz. Eq.~\ref{EOMp}, which gives $\frac{dg}{dt}=\frac{dF}{d\epsilon}\bold v\cdot (\bold{\dot k}-\bold E(t))=0$. 
Eq.~\ref{ballistic} elegantly expresses the nonequilibrium electronic distribution function $g$ as the usual Fermi-Dirac distribution with momentum shifted (minimally coupled) by an arbitrarily large impulse~\cite{mikhailov2007non}: $\bold k\rightarrow \bold k-\bold A(t) $, where  $\bold A(t)=\int^t_{-\infty}\bold E(t')dt'$. With this, the current is simply given by 
\begin{eqnarray}
\bold J[\bold A(t)]&=& \int d^3\bold k \,g(\bold k,t)\,\bold {\dot r}\notag\\
&=&\int d^3\bold k \,F\left(\epsilon\left(\bold k-\bold A(t)\right)\right) \left[\bold v-\bold\Omega\times\bold E(t) \right]\notag\\
&=& n\left[e \langle\bold v\rangle_{\bold E}+\,\frac{e^2}{\hbar}\bold E(t)\times \langle\bold \Omega\rangle_{\bold E}\right],
\label{j1}
\end{eqnarray}
where $n=\int g(\bold k,t)d^3\bold k $ is the electronic density and $\langle\bold v\rangle_{\bold E}=\int g(\bold k,t)\,\bold v \,d^3\bold k/n$, $\langle\bold \Omega\rangle_{\bold E}=\int g(\bold k,t)\,\bold \Omega \,d^3\bold k/n$ are the integrated electron velocity and Berry curvature over the Fermi region i.e. region in the Brillouin zone (BZ) where $g(\bold k,t)$ is nonzero. For small chemical potentials such that the occupied states remain close to the NLs, the Fermi region takes the shape of the nodal "tubes", but is shifted by an electromagnetic impulse $\int^t\bold E(t')dt'$ (Eq.~\ref{ballistic}).

Eq.~\ref{j1} is the main result from our semi-classical approach, and forms the starting point of our analysis~\footnote{Incidentally, it also mirrors the theoretical model for a different HHG mechanism observed in bulk ZnO~\cite{ghimire2011observation}.}. It expresses the  response current in terms of $\langle\bold v\rangle_{\bold E}$ and $\langle\bold \Omega\rangle_{\bold E}$, which themselves depend on $\bold E$ in a manner prescribed by the shape of the Fermi region. Note that both $\langle\bold v\rangle_{\bold E}$ and $\langle\bold \Omega\rangle_{\bold E}$ can both contain longitudinal and Hall components, and depend strongly on the relative signs of $\bold v$ and $\bold \Omega$ contributions within the nodal tube. Although we have derived Eq.~\ref{j1} for only the intraband contributions in the ballistic regime, our following main results, which are ultimately concerned with topological properties, should still remain valid in the presence of weak scattering.  \\

\noindent\textbf{Nonlinear response protected by nodal topology --} To illustrate how topological nodal linkages enforce nonlinearity in the current response, we first review the origin of unprotected nonlinear response in a single NL without any linkage, and next demonstrate how this nonlinearity can be protected when a nodal linkage (Hopf link) is introduced. 

\subsubsection{Single nodal loop without topological linkage}
\noindent As with other nodal systems like Graphene and Weyl semimetals~\cite{Morimoto,mooreHHG,grapheneHHG}, a single NL exhibit unconventional~\cite{lee2015negative} (though non-topological) nonlinear response behavior. With its non-topological origins, the actual extent of nonlinearity differs greatly among NLs with different dispersion and occupancy details. 
Consider a minimal NL Hamiltonian 
\medmuskip=1mu
\thickmuskip=1mu
\begin{equation}\label{H_2}
H_\text{NL}(\bold k)=\left(\cos k_x+\cos k_y+\cos k_z-r\right)\sigma_x+\sin k_z\sigma_y+M\sigma_z,
\end{equation}
where $\sigma_x$, $\sigma_y$ are the Pauli matrices, $1<r<3$ and $M$ is an extremely small gap introduced so that the Berry curvature is well-defined. To emphasize the generality of our follow arguments, we shall not fix the physical parameters till we discuss the material candidates later.
That $H_\text{NL}(\bold k)$ describes a NL can be seen from its energy dispersion
\begin{equation}\label{e_2}
\epsilon_\text{NL}(\bold k)=\pm\sqrt{M^2+\sin^2k_z+\left(\cos k_x+\cos k_y+\cos k_z-r\right)^2},
\end{equation}
which is minimally gapped along the $\cos k_x+\cos k_y=r-1$ loop in the $k_z=0$ plane. At small chemical potential $\mu> M$ and temperature $\beta^{-1}$, the occupied states are approximately contained in a tube of radius $\mu-M$ around this NL (Fig.~\ref{fig:ring}a). 

In the presence of an oscillatory electric field $\bold E(t)=E(t)\hat x$, Eq.~\ref{ballistic} dictates that the Fermi tube of occupied states is translated in the field direction according to the momentum shift $\bold k\rightarrow \bold k-\bold A(t)$. 
Due to this shift, the expectation values $\langle\bold{v}\rangle_{\bold{E}}$ and $\langle\boldsymbol{\Omega}\rangle_{\bold{E}}$ over the Fermi tube differs from their equilibrium values at $\bold E=\bold 0$, where $\langle\bold{v}\rangle_{\bold{E}=\bold 0}$ always vanishes.

As given by Eq.~\ref{j1}, the nonlinear current response arises from both the $\langle\bold{v}\rangle_{\bold{E}}$ and $\bold E\times\langle\boldsymbol{\Omega}\rangle_{\bold{E}}$ contributions. Manifestly from Fig.~\ref{fig:ring}c, the velocity field $\bold v =\boldsymbol{\nabla_p}\epsilon(\mathbf{p})$ points radially outwards outside the NL, and radially inwards inside the NL i.e. is always pointing towards higher energies. The Berry curvature $\langle\boldsymbol{\Omega}\rangle_{\bold{E}}$, on the other hand, circulates clockwise, and is concentrated along the circumference of the NL as given by
\begin{equation}\label{berry2band}
\boldsymbol{\Omega}_\text{NL}(\bold k)=\frac{M\cos{k_z}}{2|\epsilon_\text{NL}(\bold k)|^3}\begin{bmatrix}\sin{k_y}\\-\sin{k_x}\\0\\\end{bmatrix}
\end{equation}

As the applied field increases, the response behavior transits between a few distinct regimes, as illustrated in Fig.~\ref{fig:ring}c,d. In the linear regime of small $\bold A$ shifts, both $\langle\bold{v}\rangle_{\bold{E}}$ and $\langle\boldsymbol{\Omega}\rangle_{\bold{E}}$ pick up slight imbalances from their respective $\bold v$ and $\bold \Omega$ contributions, and thus rapidly increase. In particular, $\langle\boldsymbol{\Omega}\rangle_{\bold{E}}$ peaks when the clockwise/anti-clockwise vectors from $\bold \Omega$ dominate. Generically, the response has to be nonlinear due to the ``destructive interference'' of current contributions: When $\bold A$ is comparable to the NL radius, the Fermi tube straddles the interior and exterior of the NL, and receives simultaneously opposing contributions to both $\bold v$ and $\bold \Omega$. Due to the more sharply concentrated Berry flux, the $\langle\boldsymbol{\Omega}\rangle_{\bold{E}}$ contribution contains more pronounced nonlinearities. Yet, for parameters typical of nodal materials as described later, the $\langle\bold{v}\rangle_{\bold{E}}$ contribution dominates, resulting in a fairly nonlinear overall current response (Fig.~\ref{fig:ring}b).

\begin{figure*}
\begin{minipage}{\linewidth}
\includegraphics[width=\linewidth]{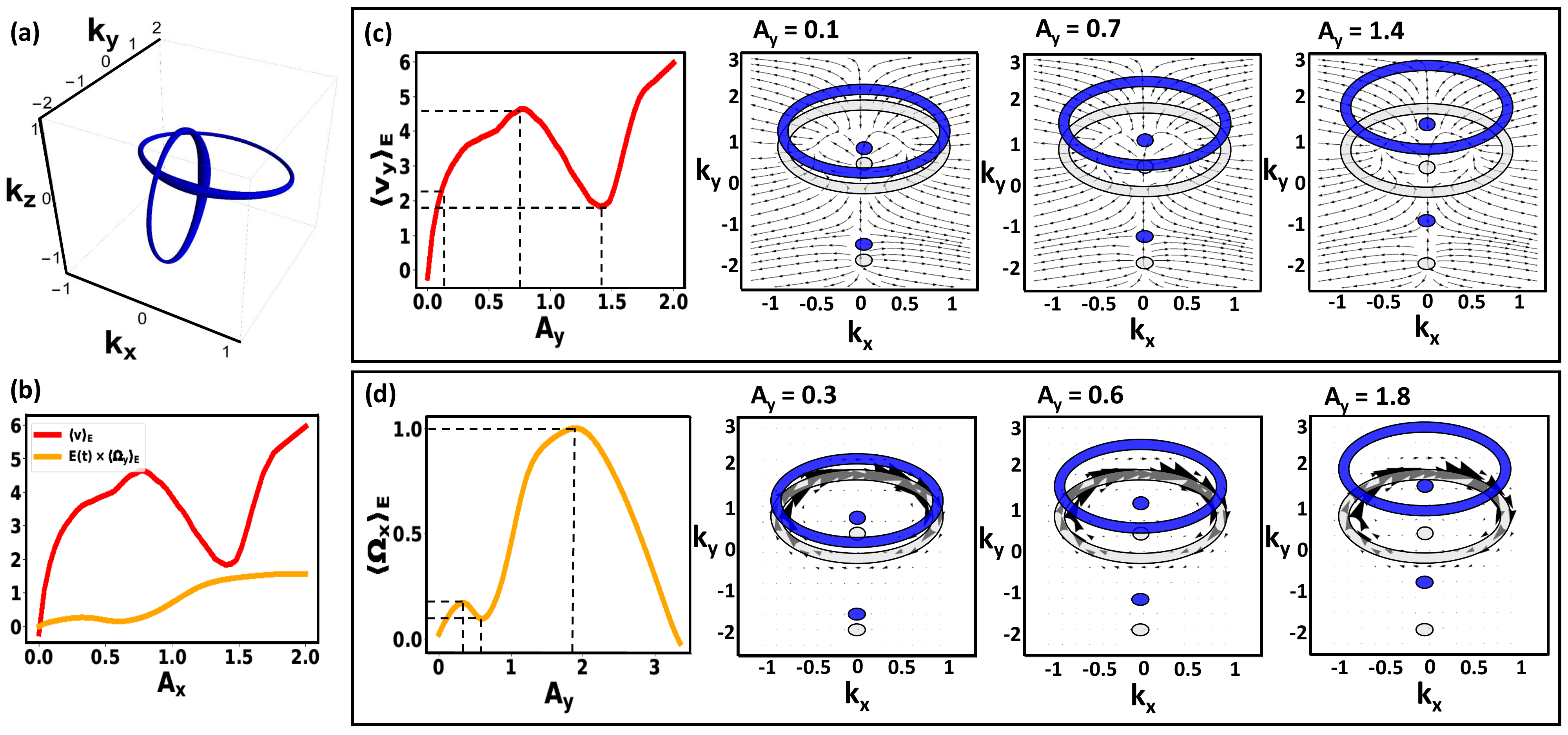}
\end{minipage}
\caption{Response nonlinearity enhanced and protected by topological linkage. (a) Nodal structure of an illustrative Hopf linkage between two NLs oriented perpendicular to $\hat e_x$ and $\hat e_z$, as defined by Eq.~\ref{hopf_H} with $\mu=0.1$, $r=2.6$ and $T=10$K. (c) $\langle v_y\rangle_{\bold E}$ contribution to the longitudinal response, which exhibits much stronger nonlinearity (with large kink) than that of the single NL of Fig.~\ref{fig:ring}. The cancellation of its $\bold v =\boldsymbol{\nabla_k}\epsilon(\mathbf{k})$ at intermediate values of $A_x$ (center) is reinforced by the singularity in $\bold v$ created by the other linked NL (small blue circular cross section). (d) $\langle \Omega_x\rangle_{\bold E}$ contribution to the transverse response, which is greatly enhanced when the other linked NL passes through the region of concentrated Berry flux without experiencing destructive interference. (b) Resultant current, which is highly nonlinear due to the topological linkage. 
}
\label{fig:Hopf}
\end{figure*}

It has to be emphasized, however, that the abovementioned response nonlinearity is not necessarily robust. In particular, destructive interference in the dominant $\langle\bold{v}\rangle_{\bold{E}}$ contribution is subject to details of the curvature in the dispersion, which should not be dependent on the nodal structure. In fact, very different response curves can result from different realizations of the same NL, as shown in the supplement~\cite{SuppMat}. As evident in Fig.~\ref{fig:ring}, the response is indeed still monotonic without a proper kink due to the ``roundness'' of the dispersion at the center of the NL.


\subsubsection{Nodal loops with topological linkage}

\noindent To ensure pronounced nonlinearity in the current response $\bold J[\bold A(t)]$, significant destructive interference of $\langle\bold{v}\rangle_{\bold{E}}$ must be guaranteed at certain values of the applied field $\bold A$. We shall see that this is assured if the NL is linked by another NL, such that an unremovable singularity passes through it. Consider a simplest nodal Hopf link Hamiltonian~\cite{lee2019imaging} possessing two interlinked NLs (Fig.~\ref{fig:Hopf}):
\begin{equation}\label{hopf_H}
H_\text{Hopf}(\mathbf{k})=Re[f(v,w)]\sigma_x+Im[f(v,w)]\sigma_y+M\sigma_z
\end{equation}
where $M$ is a small gap and $f(v,w)=(v-w)(v+w)$ with $v=\sin
(k_z-k_x)+i(2\cos k_x\cos k_z+\cos k_y-r)$ and $w=\sin(k_x+k_z)+i\sin k_y$. $1<r<3$ controls the shape and size of each loop as in the NL model, as well as their relative separation. As shown in Fig.~\ref{fig:Hopf}a, it consists of two linked NLs normal to $\hat x$ and $\hat z$, with dispersion explicitly given by
\begin{equation}\label{hopf_energies}
\epsilon_\text{Hopf}(\bold k)=\pm\sqrt{M^2+\chi^2+4(e_1e_2-e_3e_4)^2},
\end{equation}
where substitutions $\chi\equiv e_1^2-e_2^2-e_3^2+e_4^2$, $e_1\equiv\sin(k_z-k_x)$, $e_2\equiv(\cos(k_x+k_z)+\cos{k_y}+\cos(k_z-k_x)-r)$, $e_3\equiv\sin(k_x+k_z)$ and $e_4\equiv\sin{k_y}$ have been defined for notational brevity.  As plotted in Fig.~\ref{fig:Hopf}b, it gives rise to a $\bold v=\boldsymbol{\nabla_k}\epsilon(\mathbf{k})$ field with sources or sinks within the NLs. 
These singularities are the cross sections of the other topologically linked NL, which is analytically given by simultaneous solutions to $e_1^2+e_4^2=e_2^2+e_3^2$ and $e_1e_2=e_3e_4$. 

We next show that singularities of the $\bold v$ vector field can indeed ensure significant nonlinearity in the current response $\bold J(\bold A)$ by studying the differential response of a thin NL (derived in the Supplement):
\begin{eqnarray}
\frac{d J_{i}}{d  A_{j}}&\approx &\, 2 \sum_{\alpha\in \text{NLs}}\frac{\mu}{\hat e_j\cdot\bold v_F}\frac{d^2 \epsilon(\vec k^\alpha+\vec A)}{dk_i dk_j}.
\label{dJdp}
\end{eqnarray}
In the above, $\vec k^\alpha$ labels the trajectories of all the NLs in the BZ, and $\hat e_j\cdot \bold v_F$ is the component of the Fermi velocity of the $\alpha$-th NL at momentum $\vec k^\alpha$ along the applied field. In particular, the longitudinal differential response is proportional to the sum of the curvature of the dispersion along the $\vec A$-displaced NLs. When the curvatures conspire to form a highly fluctuating sum, the current response contains large fluctuating gradients, and becomes significantly nonlinear.

For concreteness, we analyze Eq.~\ref{dJdp} with the simplest possible in-plane dispersion hosting a nodal linkage: $\epsilon_\text{link}^0(k_x,k_y)= \left|\sqrt{k_x^2+k_y^2}-r\right|$, which describes a circular NL of energy $\epsilon=0$ and radius $r$ in the $k_x$-$k_y$ plane, which is perpendicularly linked by another NL of energy $\epsilon=r$ at $\bold k=\bold 0$. To regularize the infinite dispersion curvature at its nodes, we introduce small gaps~\footnote{Qualitatively, these gaps can represent the smudging of the Fermi tube due to nonzero temperature.} $m_0$, $m_r$ at the perpendicular and in-plane rings at $|\bold k|=0$ and $r$, such that the dispersion is modified to 
\begin{eqnarray}
\epsilon_\text{link}(k_x,k_y)&=& \sqrt{m_r^2+\left(\sqrt{k_x^2+k_y^2+m_0^2}-\sqrt{r^2+m_0^2}\right)^2}\qquad.
\label{diracE}
\end{eqnarray}
Its second derivative $\frac{d^2\epsilon_\text{link}}{d^2k_j}$, $j=x,y$ almost vanishes except at the nodes, where it diverges like $\sim -m_0^{-1}$ at the nodal linkage through the origin, and like $\sim m_r^{-1}k_j^2r^{-2}$ on the $k_x^2+k_y^2=r^2$ nodal ring. These divergences are exactly the large curvature fluctuations necessary for a greatly fluctuating differential longitudinal response $\frac{dJ_j}{dA_j}$ (Eq.~\ref{dJdp}). When a (regularized) nodal linkage is present, small gaps $m_0,m_r$ and thus large curvature fluctuations leading to response nonlinearity are inevitable.

We next analyze the response kinks corresponding to these curvature fluctuations. At very small $\bold A_j$ where the Fermi tube is almost aligned with the NL, $\frac{d^2\epsilon_\text{link}(\bold k^\alpha+\bold A_j)}{d^2k_j}$ picks up large positive contributions $\sim m_r^{-1}k_j^2r^{-2}$, giving rise to large differential response. But at $\bold A_j\approx r$, $\frac{d^2\epsilon_\text{link}(\bold k^\alpha+\bold A_j)}{d^2k_j}$ picks up a large negative contribution $~m_0^{-1}$ as it cuts the origin, and hence produces a negative differential response. In fact, the differential response of this model has to exhibit a \emph{reversal} due to the opposite signs of dominant curvature contributions, thereby leading to a very \emph{non-monotonic} and hence highly nonlinear current response containing kink/s. In generic topologically linked NL systems, the linkages still enforce similar oppositely signed dispersion curvature contributions, and it is in this sense that their $\langle\bold{v}\rangle_{\bold{E}}$ response nonlinearity is protected. Strong non-linearities also analogously dominate the $\langle\bold{\Omega}\rangle_{\bold{E}}$  contribution to the response, but they typically play a subdominant role. 

\begin{figure}
\begin{minipage}{\linewidth}
\includegraphics[width=\linewidth]{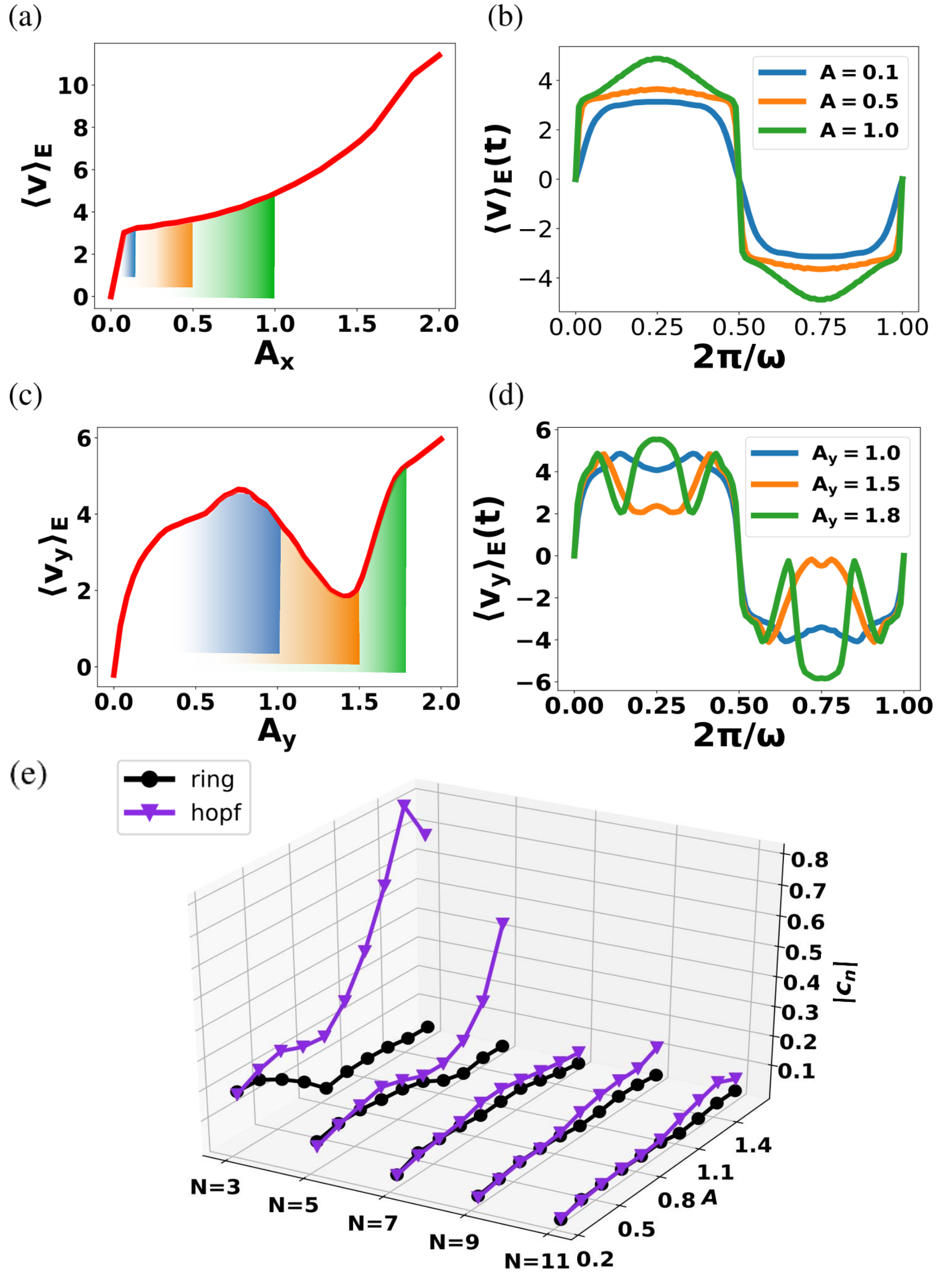}
\end{minipage}
\caption{(a,c) Response curves of the single NL and Hopf nodal systems of Figs.~1 and 2 respectively, with (b,d) demonstrating how they distort sinusoidal signals of different amplitudes $A$ (as indicated by corresponding colored regions in the response curves). Due to the non-monotonic response in the topologically linked Hopf system (c), signal oscillations traversing the kink acquire additional fluctuations of higher frequency (d). The superior HHG by the Hopf system is evident from its larger Fourier coefficient ratios $c_n$ (Eq.~\ref{Jj})  in (e).
}
\label{fig:HHG}
\end{figure}

\begin{figure*}
	\subfloat[]{
		\includegraphics[width=.15\linewidth]{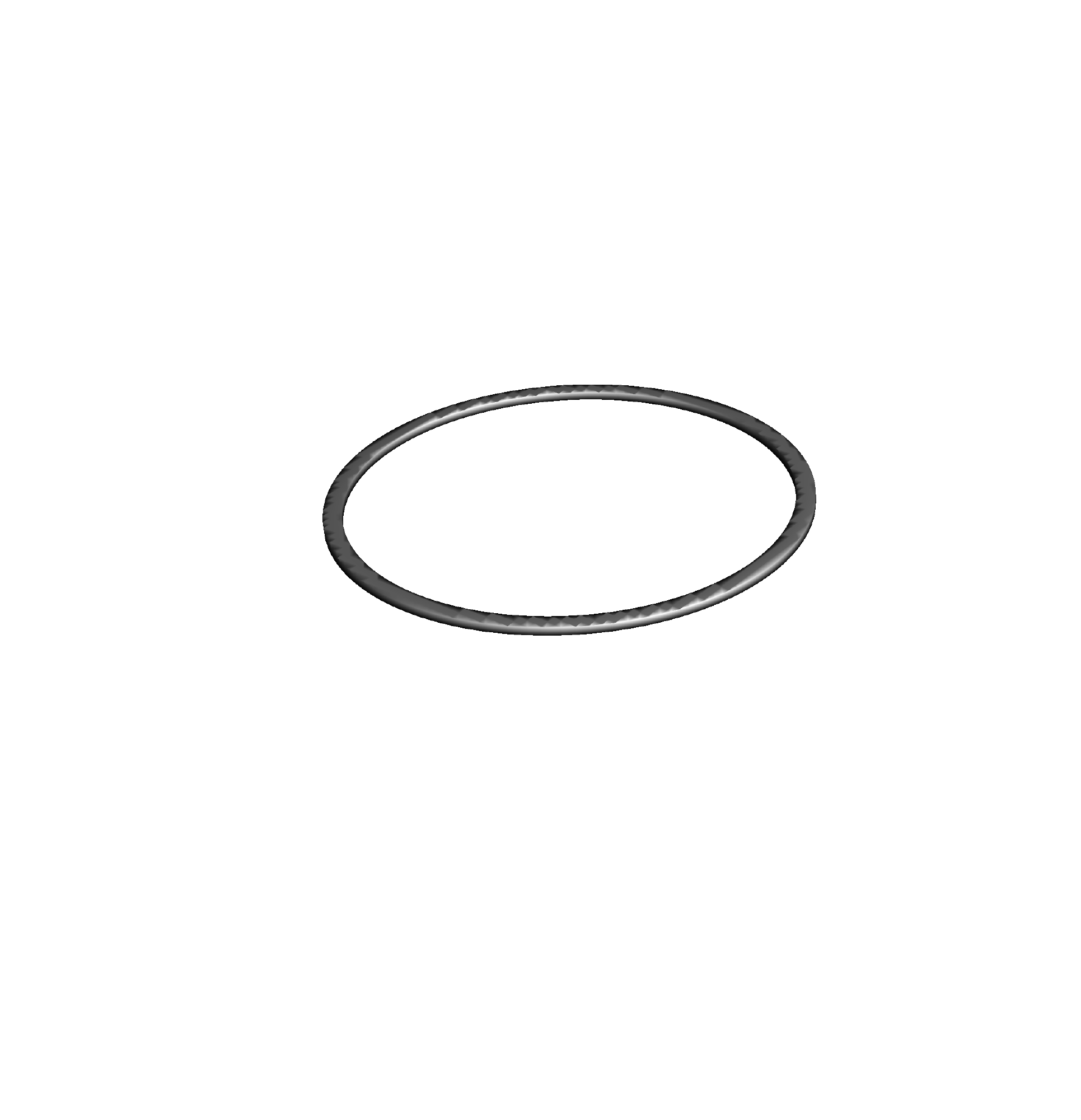}
	}
	\subfloat[]{
		\includegraphics[width=.15\linewidth]{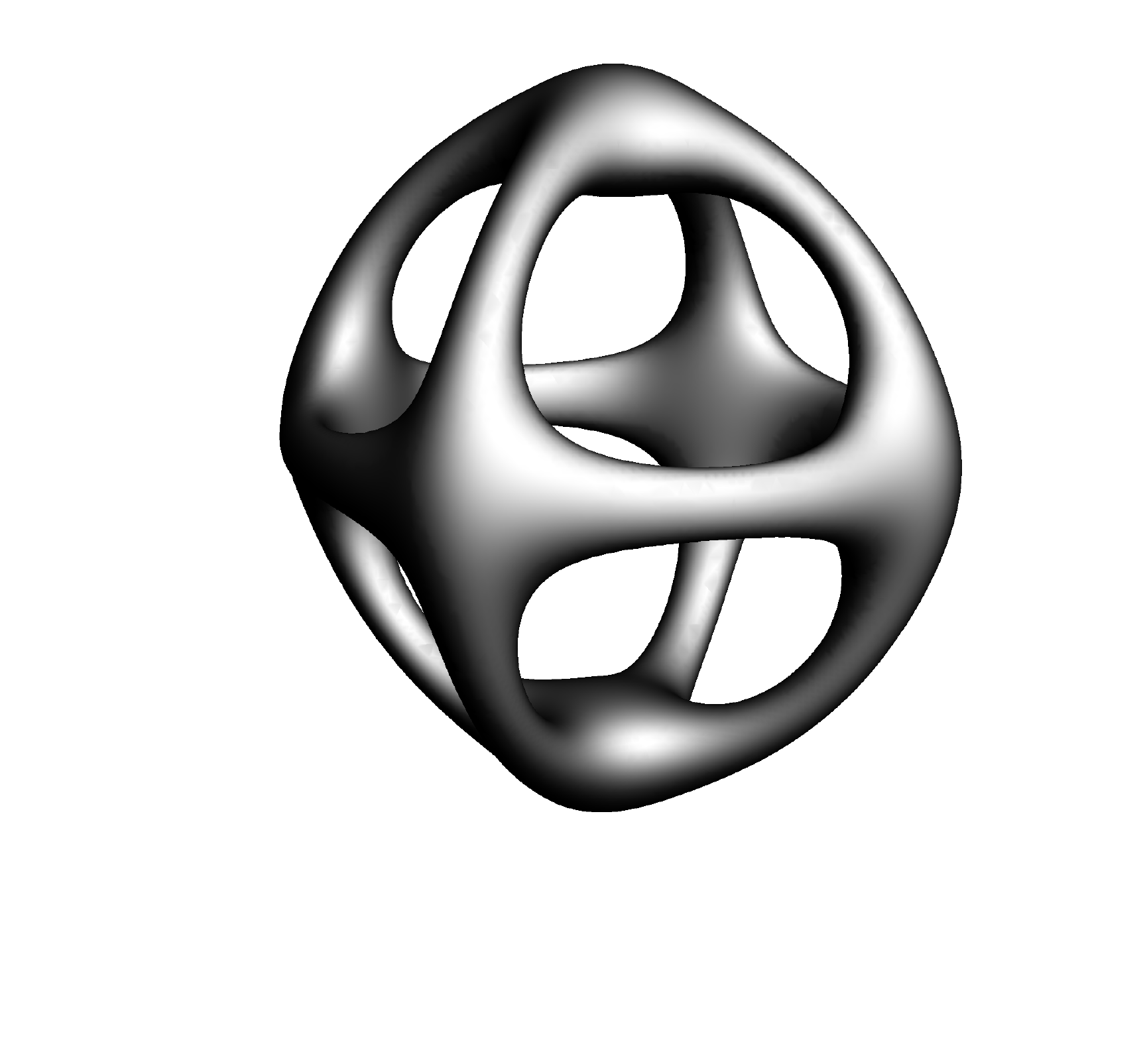}
	}
	\subfloat[]{
		\includegraphics[width=.17\linewidth]{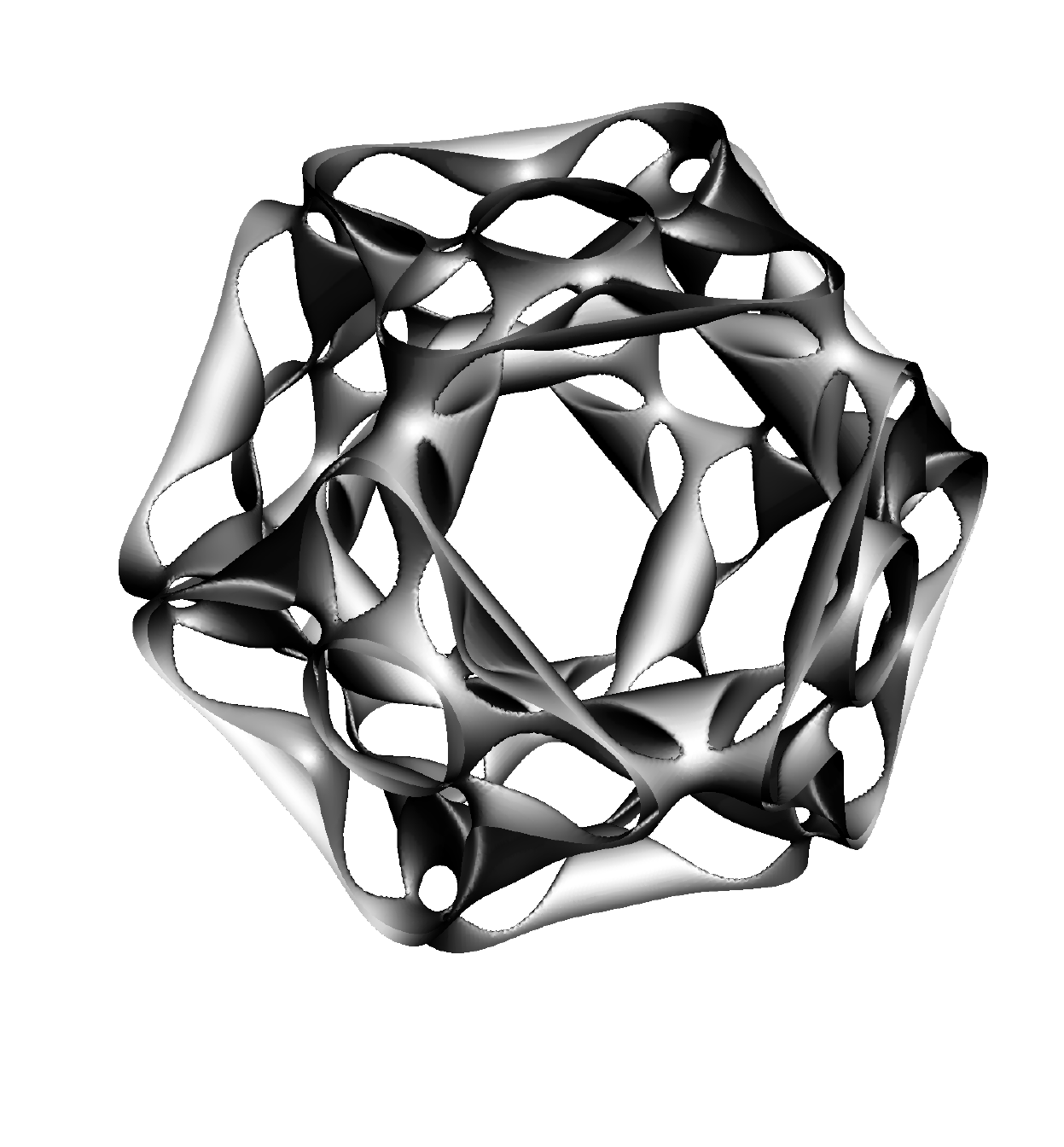}
	}
	\subfloat[]{
		\includegraphics[width=.24\linewidth]{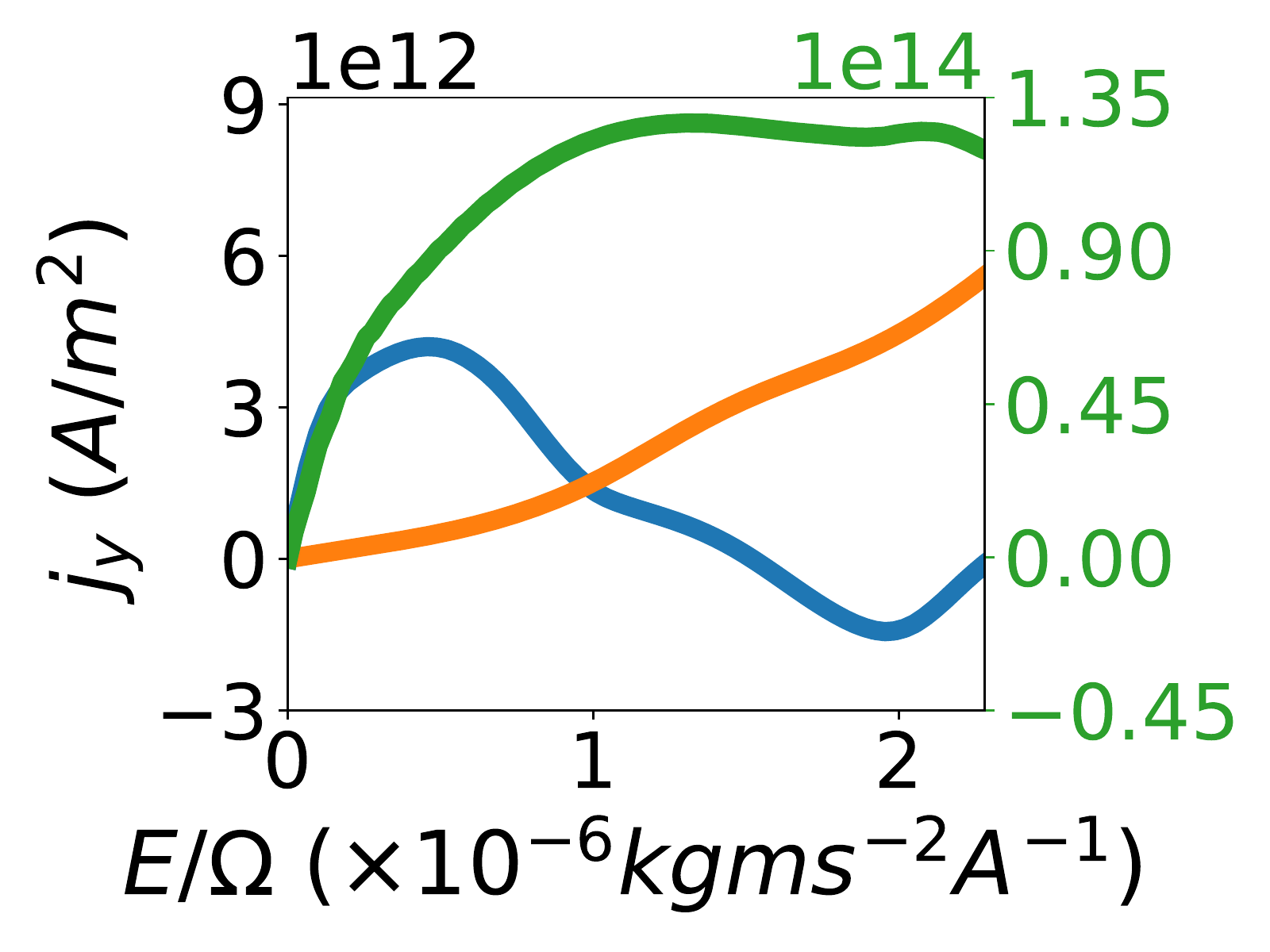}
	}
	\subfloat[]{
		\includegraphics[width=.27\linewidth]{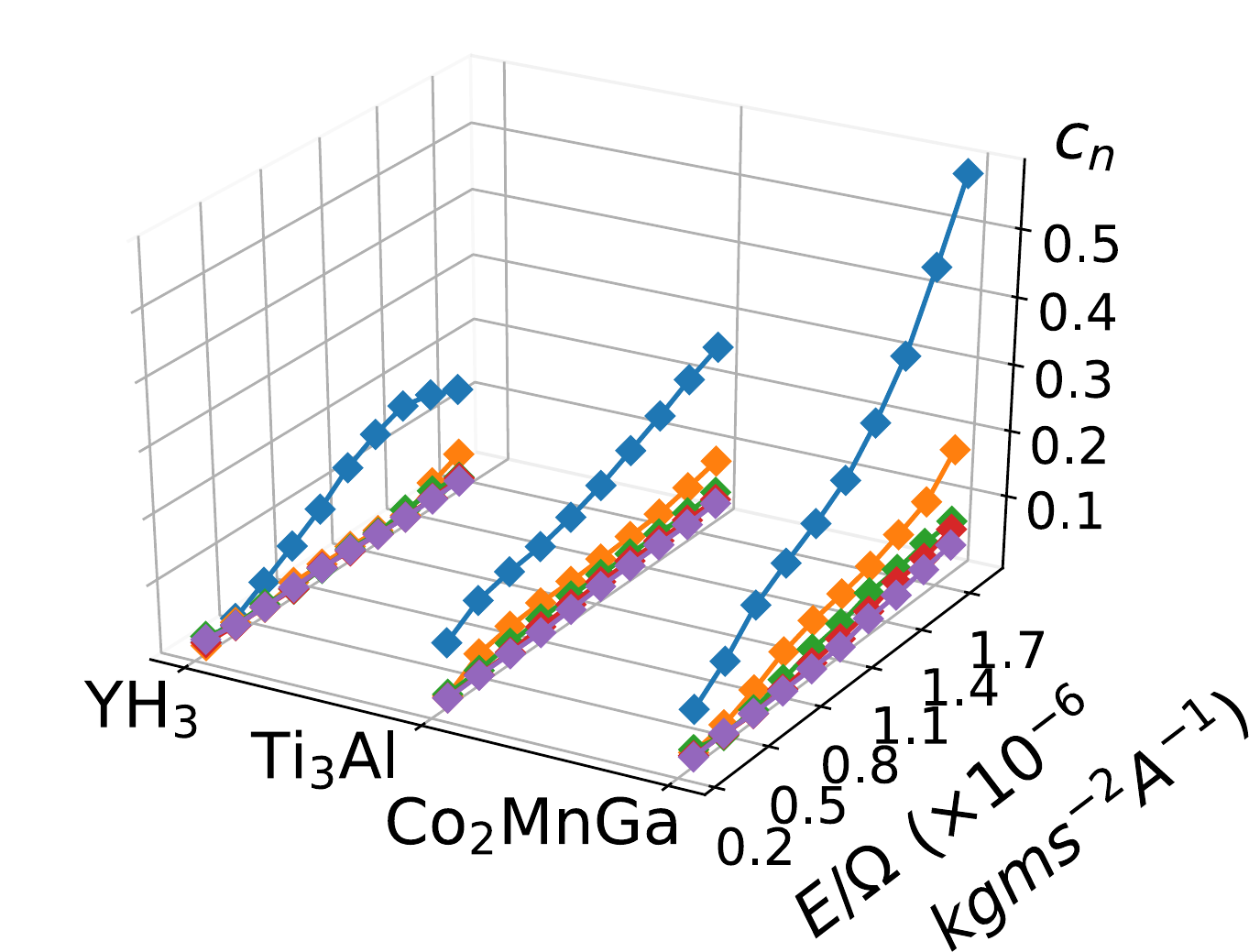}
	}	
	\caption{Occupied states (Fermi seas) of (a) Ti$_3$Al, (b) YH$_3$ and (c) \CMG. (d): The current response of  Ti$_3$Al (green), YH$_3$ (orange) and Co$_2$MnGa (blue) at 10 K with chemical potentials $\mu=0.15$ eV, $0.002$ eV and $0.130$ eV respectively. The left and right vertical axes denote the numerical current in YH$_3$/Co$_2$MnGa and Ti$_3$Al, arising from electromagnetic impulses $E/\Omega$. (e): The higher harmonic ratios $c_n$ as a function of the driving impulse amplitude $p_0$, with Co$_2$MnGa exhibiting by far the strongest HHG due to its multiple nodal linkages. For clarity, the HHG ratios are also individually plotted out in the appendix.}
	\label{fig:materials}
\end{figure*}

\noindent\textbf{Higher harmonic generation from nonlinear response --} Consider a sinusoidal time-varying applied electric field signal $\bold E(t)= \bold E_0\cos\Omega t$, which corresponds to the vector potential $\bold A(t)=\bold A\sin\Omega t$, where $\bold A =\frac{\bold E_0}{\Omega}= \frac{\bold p_0a}{\hbar}$, $\bold p_0$ the impulse amplitude vector and $a$ the lattice spacing of the nodal material lattice. We have temporarily reintroduced the physical units in preparation for later experimental discussion. The greater the nonlinearity of the current response function $\bold J(\bold A)$, the larger is the extent of HHG. The extent of signal distortion between chosen direction components $i$ and $j$ can be quantified via Fourier coefficient ratios $c_n$ describing the higher harmonics:
\begin{eqnarray}
J_j(A_i(t))&=&J_j\left(A\sin\Omega t\right)\notag\\
&\propto& \,\sin \Omega t +\sum_{n\neq 0}c_n \sin n\Omega t.
\label{Jj}
\end{eqnarray}
In Fig.~\ref{fig:HHG}c,d, we see that the pronounced non-monotonicity of the $\langle\bold v\rangle_{\bold E}$ part of the Hopf current response leads to the creation of new higher frequency peaks. While higher harmonics are also present in the the single NL response (Fig.~\ref{fig:HHG}a,b), they are manifestly smaller, corresponding only to gentle corrections to sinusoidal current output. As summarized in Fig.~\ref{fig:HHG}e, the Hopf linkage indeed possess much larger $c_n$ for most amplitudes, especially when the amplitude corresponds to the response kink where $c_3$ and $c_5$ are close to unity. In more generic NL configurations i.e. a Borromean Ring with topological linkages in multiple planes, the response function in the direction parallel to each plane will necessarily all exhibit large nonlinearity and hence strong HHG. 

The leading nonlinear response is also quantified by the 3HHG and Kerr coefficients. Expanding Eq.~\ref{Jj} up to the third order and considering only the longitudinal components for simplicity, we can write the response current in a chosen direction as $J=\alpha A + \beta A^3$. The third order response, which is proportional to $\beta$, can be further classified into the 3HHG and Kerr contributions with frequencies $3\Omega$ and $\Omega$ respectively. For instance, the material Co$_2$MnGa introduced below possesses 3HHG and Kerr coefficients respectively given by  $(7.75+0.82i)\times 10^{-10}s^9A^4kg^{-3}m^{-5}$ and $(2.32+0.24i)\times 10^{-9}s^9A^4kg^{-3}m^{-5}$ at $\Omega=16$ THz, $\tau=10^{-13}$s, $\mu=0.15$eV and room temperature $300$K.

\noindent\textbf{HHG in nodal material candidates -- } We now present quantitative calculations of the HHG induced by the nodal topology of three different known nodal materials, arranged in order of increasing topological sophistication. We shall be exclusively concerned with the $\langle \bold v\rangle_{\bold E}$ response, since the Berry curvature contribution is typically smaller and depends non-universally on the internal structure of the eigenbands.

\subsubsection{1. Ti$_3$Al - single nodal ring}

One first nodal material is Ti$_3$Al, which contains a single almost ``ideal" NL whose band crossing is almost dispersionless and very close to the Fermi level~\cite{zhang2018nodal}. These properties yield a well-defined nodal ring with almost uniform thickness (Fig.~4a), qualitatively approximating our toy model from Eq.~\ref{e_2}. While its nodal structure has yet to be experimentally verified, high quality samples of Ti$_3$Al already exist in both alloy and nanoparticle form~\cite{basuki2016interdiffusion,yang2003phase,chen2013bulk}.

To compute its $\langle \bold v\rangle_{\bold E}$ current response, we employ its DFT-fitted energy dispersion from Ref.~\cite{zhang2018nodal}, as described in the Methods~\cite{SuppMat}. 
Like the single NL toy model, Ti$_3$Al exhibits some nonlinearity in its current response, although hardly large enough to exhibit non-monotonicity (green curve in Fig.~4d). Correspondingly, its HHG ability is limited, with even the lowest $n=3$ harmonic generation ratio $c_3$ being below $30\%$ (Fig.~4e). 

\subsubsection{2. YH$_3$ - three touching nodal rings}
Next up in sophistication is the material YH$_3$, which is proposed to contain three orthogonal NLs that surround the $\Gamma$-point~\cite{shao2018nonsymmorphic}. These NLs are protected by non-symmorphic symmetry present in its P$\bar 3$c1 form, which likely to be its most stable crystal structure as seen from neutron diffraction experiments~\cite{fedotov2006displacive,udovic2008nature}.

While its three NLs are isolated from other bands, they intersect instead of linking each other (Fig.~4b). To explore whether the absence of linkages penalize HHG, we first construct its tight-binding model according to a DFT-fitted ansatz detailed in the Methods~\cite{SuppMat}.
As evident in Figs.~4d,e, the resultant $\langle \bold v\rangle_{\bold E}$ response non-linearity and hence HHG for the three touching NLs of YH$_3$ is even weaker than that of the single NL Ti$_3$Al, even before spin-orbit coupling effects gap out NLs into elogated Fermi tubes. The reason for the poor HHG is that in this case, the multiple NLs ``smooth'' out the energy dispersion around the nodal touchings, thereby reducing the total potential for destructive interference of velocity velocities.  

\begin{figure*}
	\begin{minipage}{\linewidth}
		\includegraphics[width=\hsize]{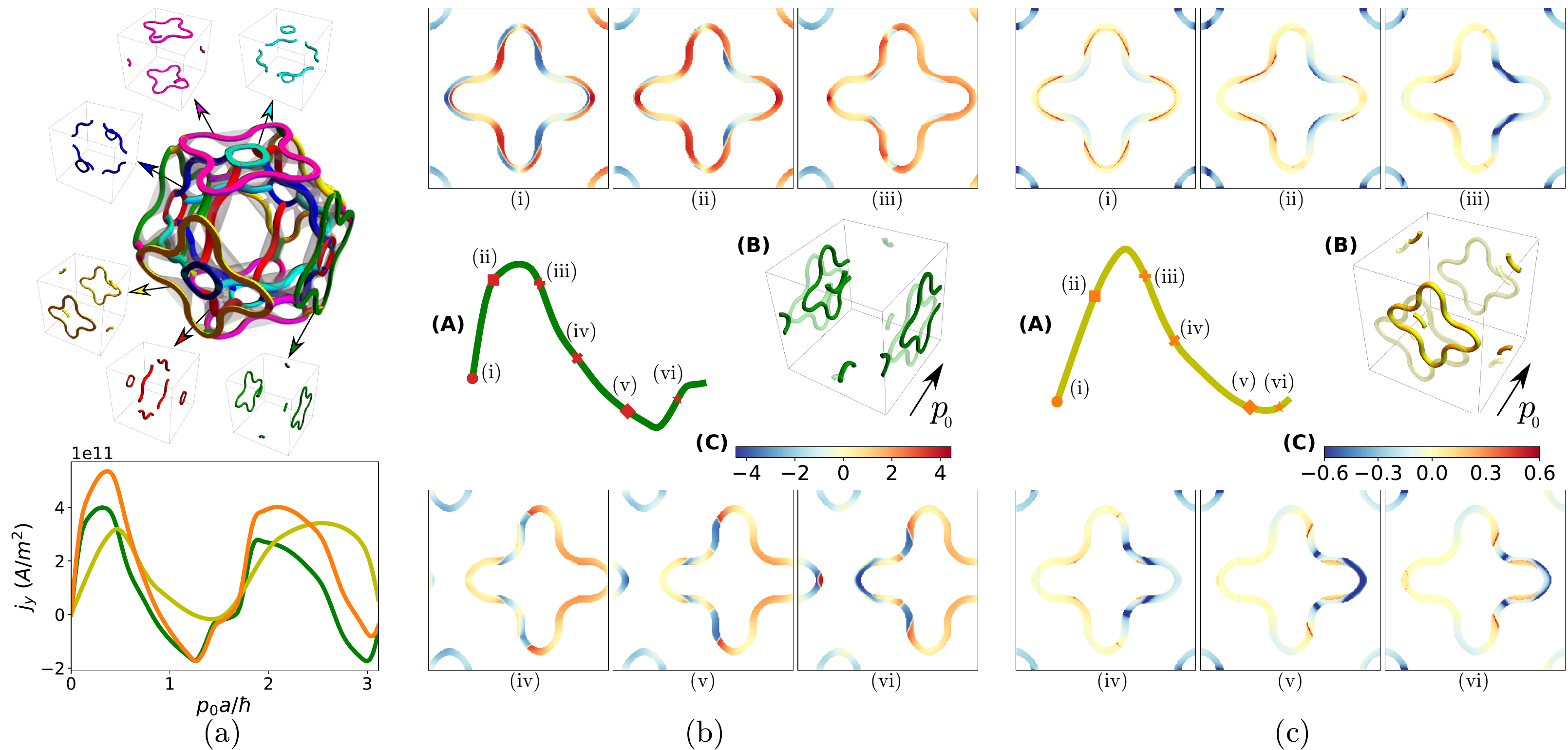}
	\end{minipage}
	\caption{Decomposition of the current response of Co$_2$MnGa into those of its constituent nodal features. (a) Top: Breakdown of its nodal structure into dominant nodal Hopf flower petals and rings whose linkages are further detailed in Fig.~6. Bottom: Total current response (orange, scaled down by a factor of four for clarity) from the sum of green/magenta/red/cyan and yellow/blue nodal structure responses, which correspond to only two unique curves due to cubic symmetry. (b,c): (A) response curves of the green/yellow Hopf flower petals in the directions indicated by the impulse $p_0$, as illustrated in the insets (B). The response curve shapes at various impulse strengths (i-vi) can be understood through integrating the explicit electronic group velocity profiles within the petals, as illustrated in the corresponding upper and lower panels. Legend colorbars (C) for the group velocities are given in units of $\times 10^{5}$ ms$^{-1}$. 
		}
	\label{fig:CMGmain}
\end{figure*}

\subsubsection{3. Co$_2$MnGa - intertwined nodal flower petals, rings etc.}

We now turn to our third and most topologically sophisticated material, the magnetic Heusler compound Co$_2$MnGa which was theoretically predicted in 2017~\cite{CMG_theory} and soon realized experimentally by various groups~\cite{CMG_ARPES,CMG_Nernst,CMG_AHE}. In Ref.~\cite{CMG_ARPES}, the intricate nodal structure of Co$_2$MnGa was revealed to contain Hopf links, inner and outer chains at the Fermi energy, as shown in Fig.~4(c) and~\ref{fig:CMGmain}. While Co$_2$MnGa contains even more nodal components than YH$_3$, the crucial difference is that many of its components are \emph{topologically linked}, and not just touching each other.

\begin{figure*}
	\begin{minipage}[b]{\textwidth}
		\subfloat[]{\includegraphics[width=.2\linewidth]{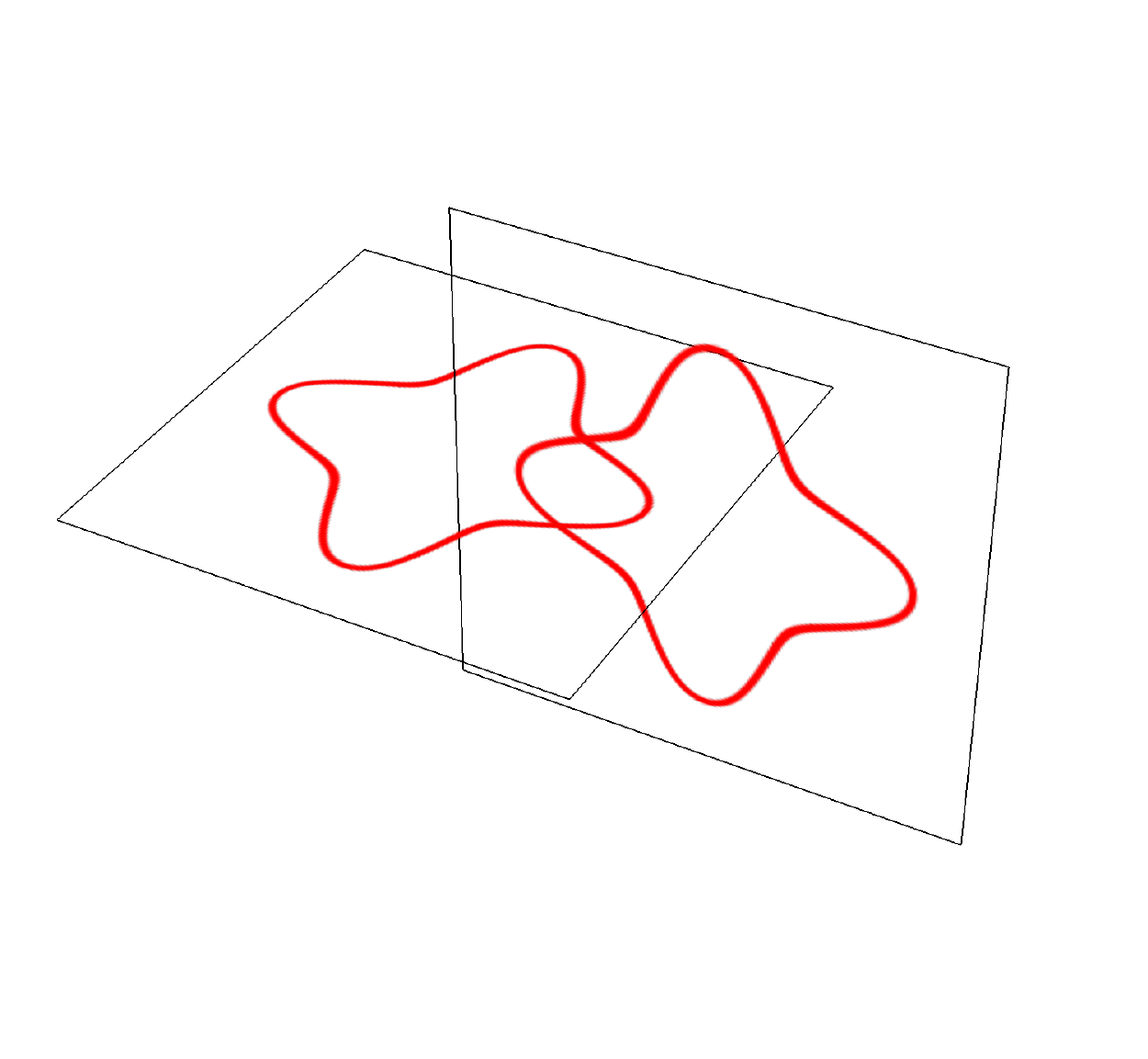}}
		\subfloat[]{\includegraphics[width=.2\linewidth]{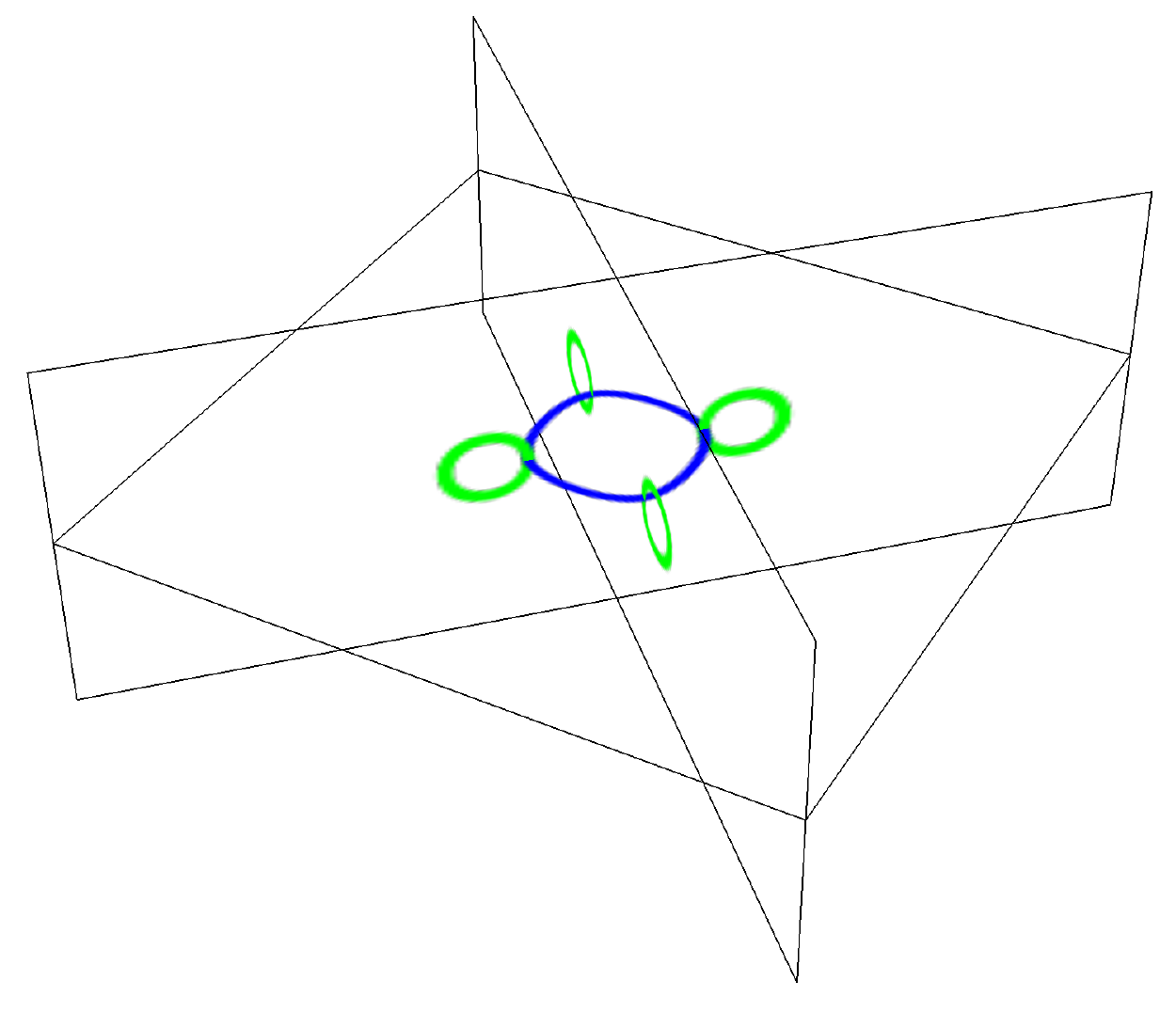}}
		\subfloat[]{\includegraphics[width=.2\linewidth]{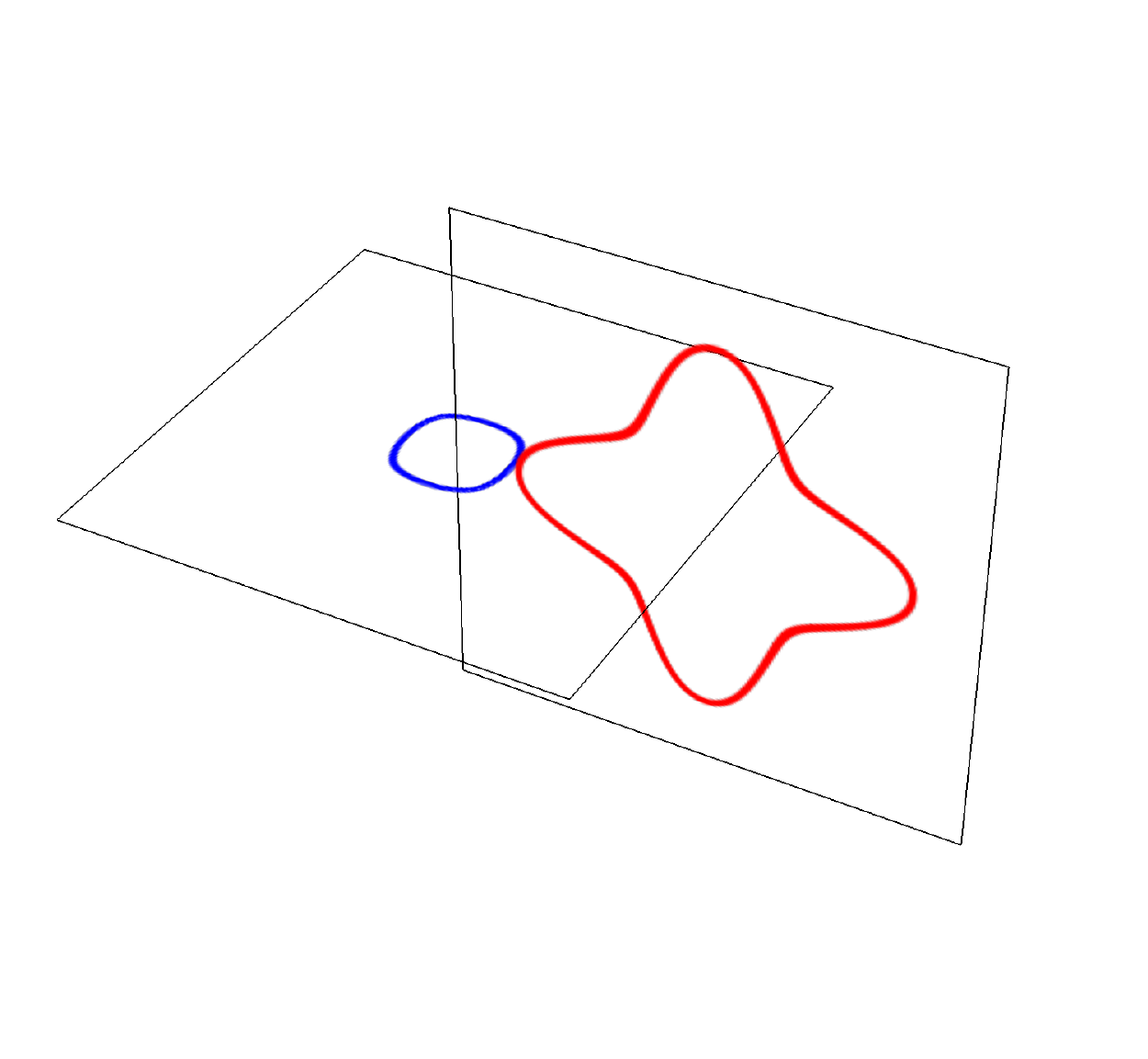}}
		\subfloat[]{\includegraphics[width=.2\linewidth]{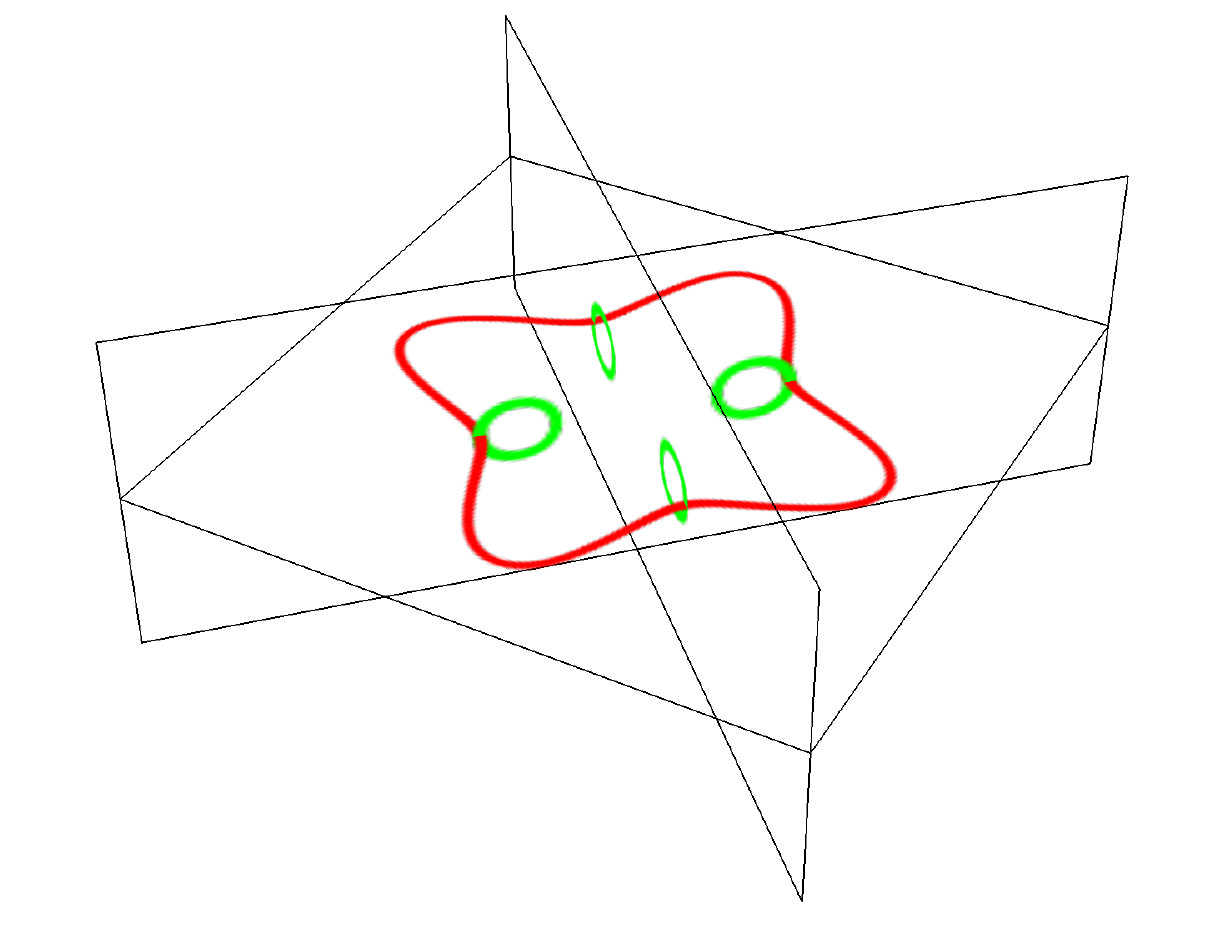}} 
		\subfloat[]{\includegraphics[width=.2\linewidth]{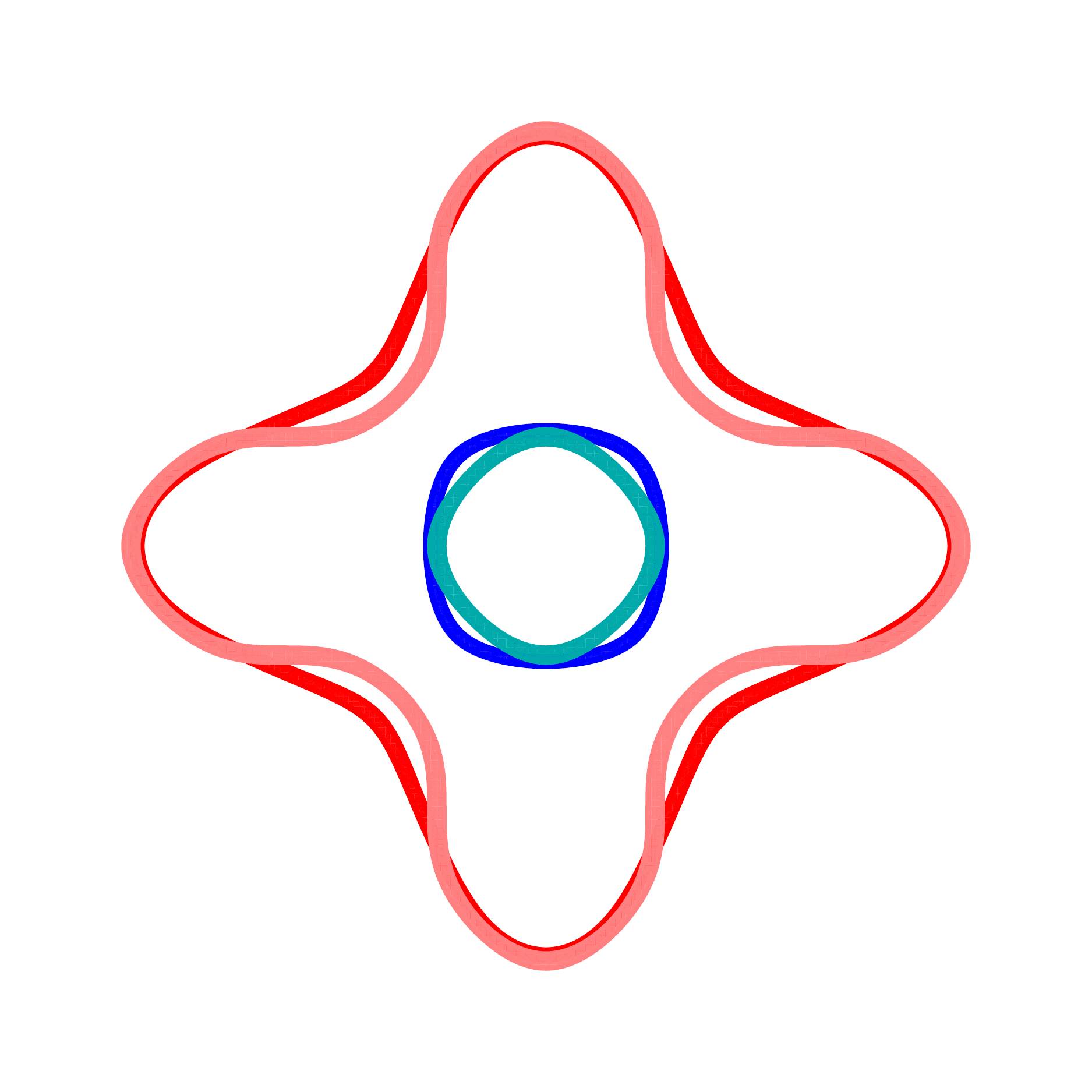}} 
		\\
		\subfloat[]{\includegraphics[width=.2\linewidth]{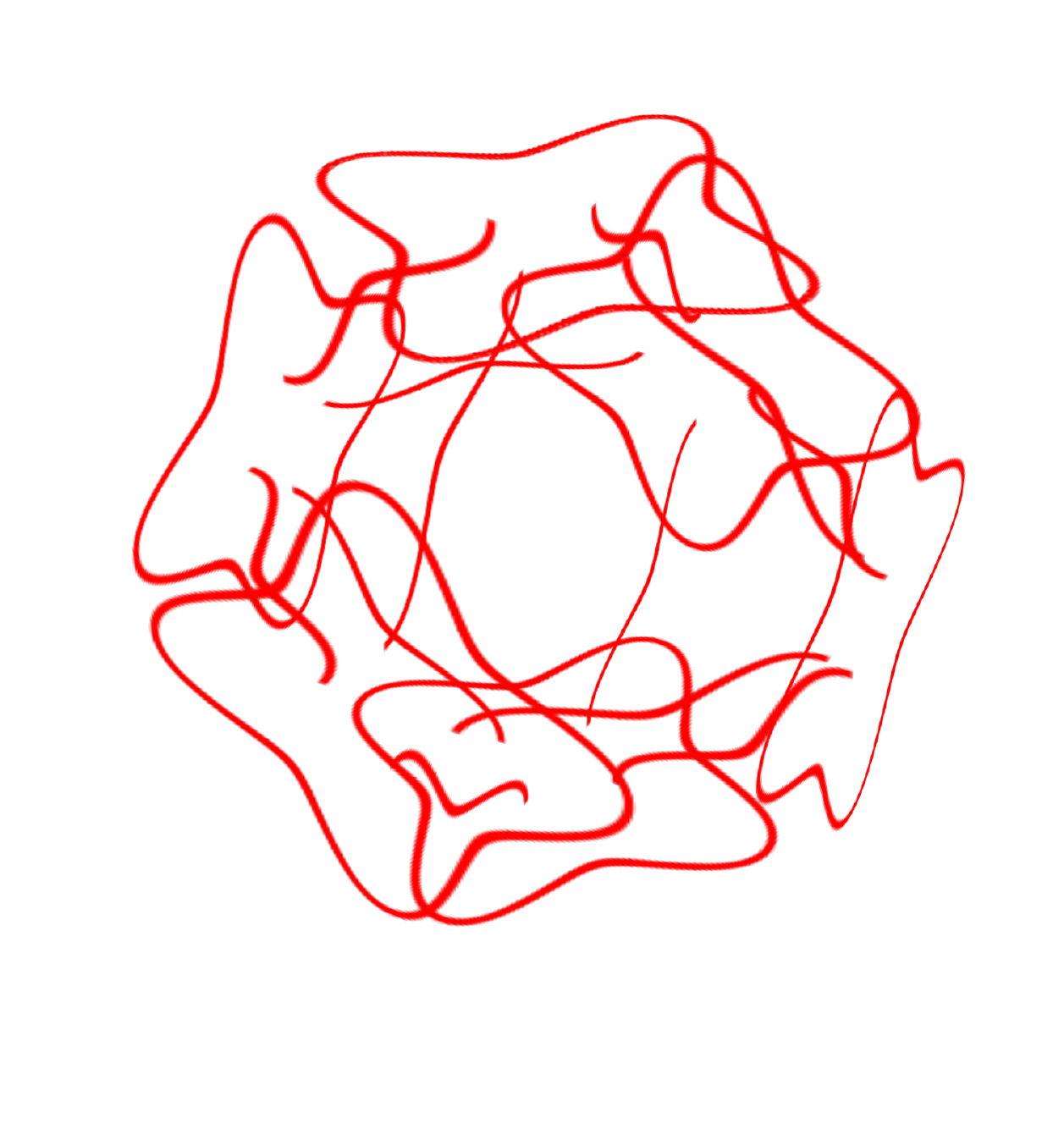}}
		\subfloat[]{\includegraphics[width=.2\linewidth]{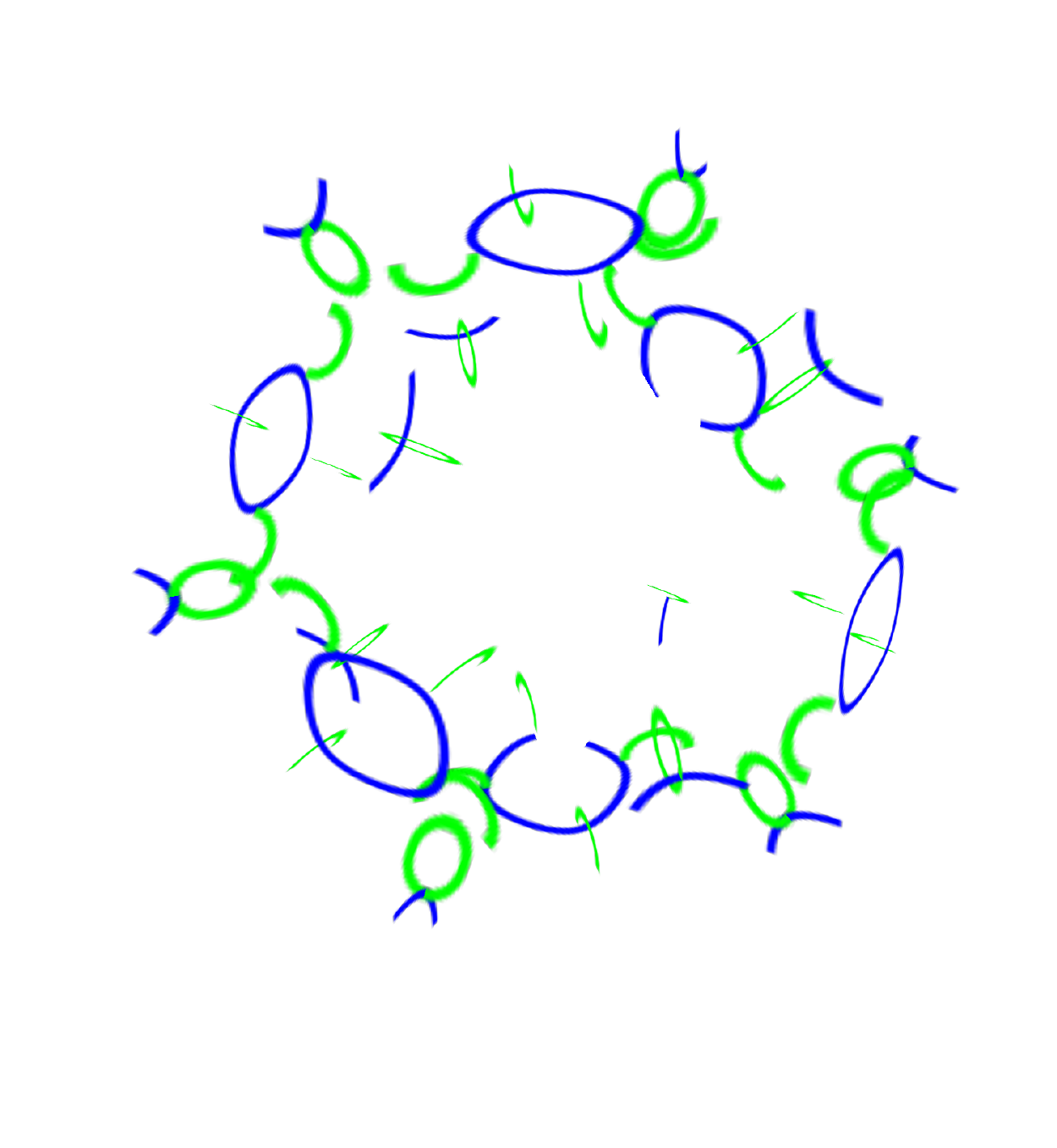}}
		\subfloat[]{\includegraphics[width=.2\linewidth]{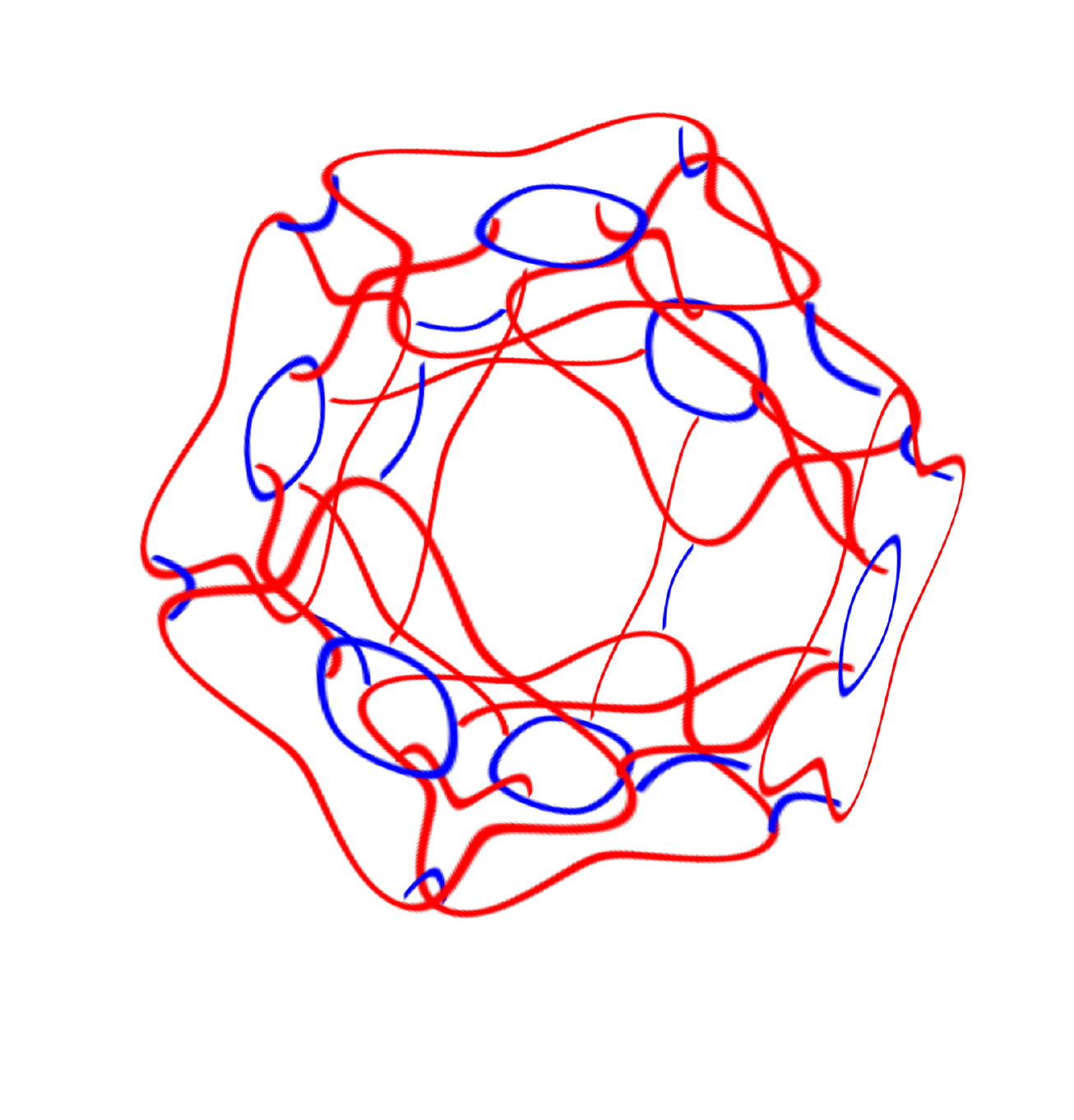}}
		\subfloat[]{\includegraphics[width=.2\linewidth]{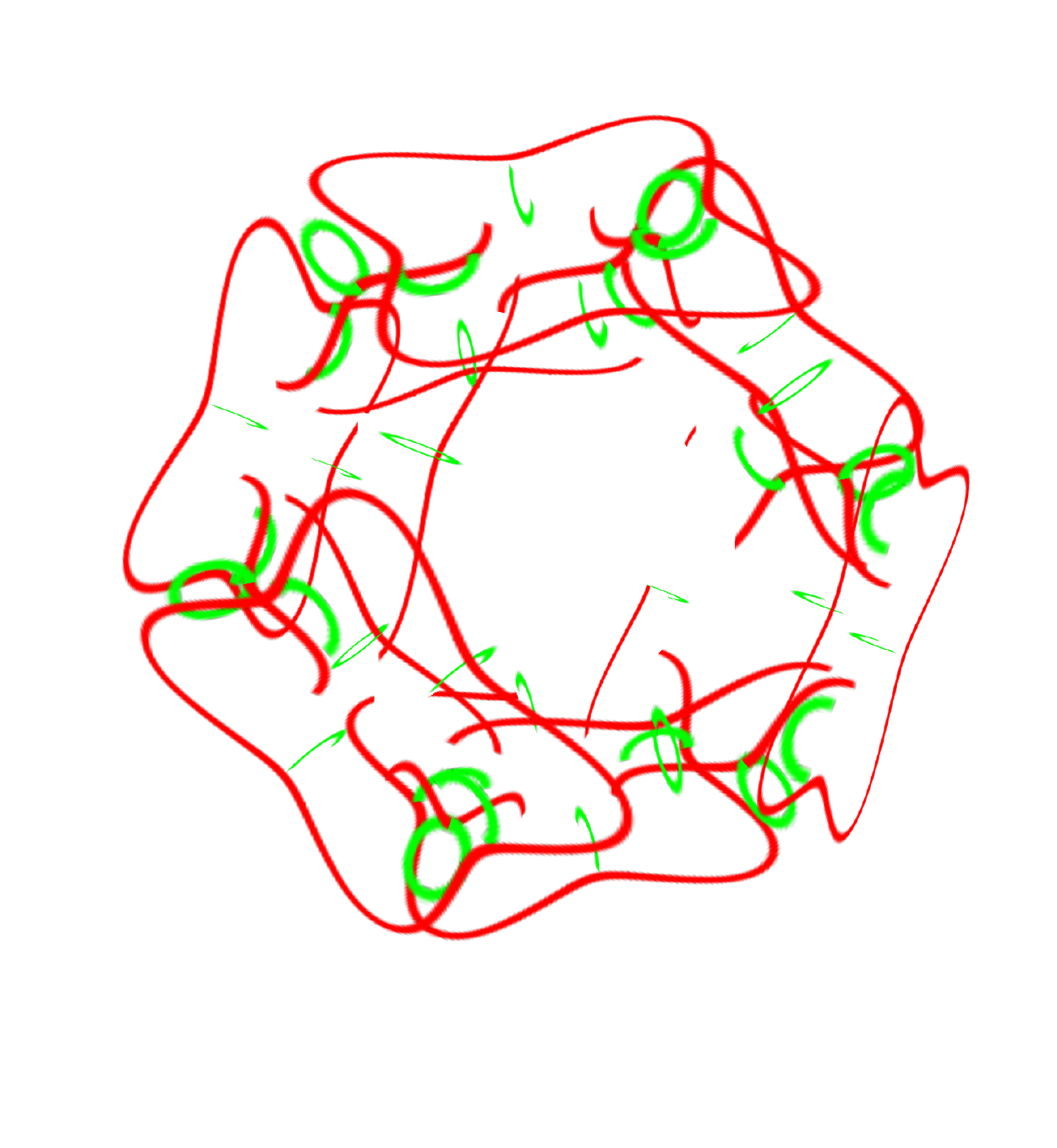}}
		\subfloat[]{\includegraphics[width=.2\linewidth]{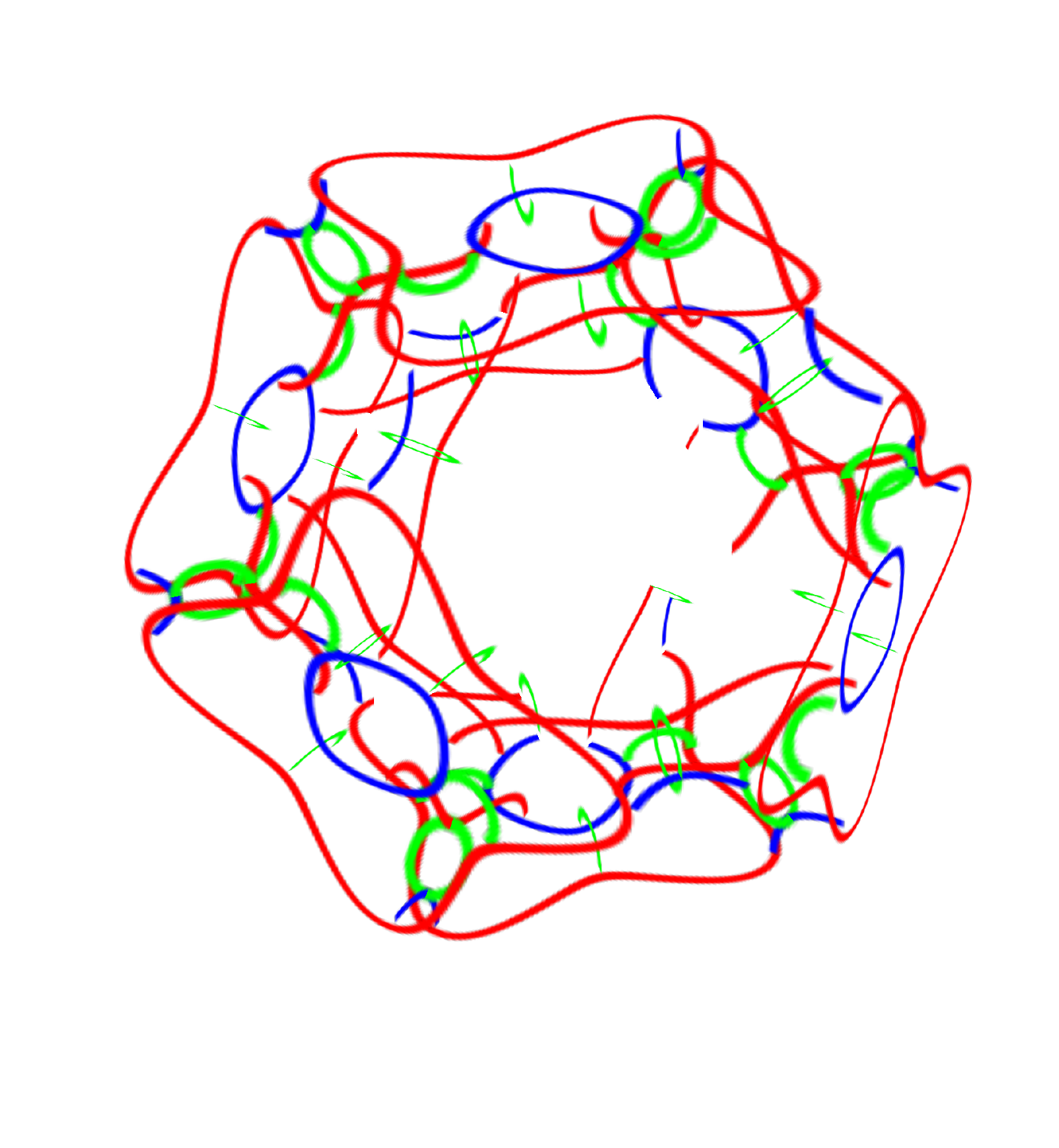}}
	\end{minipage}
	\caption{The assembly of all the nodal features of CoMn$_2$Ga into a linked nodal network. Mostly saliently, it involves Hopf links [Fig.~\ref{fig:CMG}(a),(f)], inner chains [Fig.~\ref{fig:CMG}(d),(i)] and outer chains [Fig.~\ref{fig:CMG}(b-c),(g-h)]. Specifically, it consists of a (a) Hopf link between red flower petals, (b) outer chain involving a blue loop and four smaller green loops, (c) outer chain formed by a blue loop and a flower petal, (d) inner chain due to a flower petal and four green rings. These features form sub-networks of linkages: (f) Network of Hopf links, (g) network of outer chains formed from blue and green loops, (h) network of outer chains formed from blue loops and red flower petals, (i) network of inner chains, which together form the full nodal network (j). Within it, the red flower petals and blue loops occupies the greatest volume and hence dominate the response. As shown in (e), their Fermi surfaces are decently approximated by our two-band model (pink flower petal and cyan loop). For graphical completeness, all nodal structures are displayed not in the first Brillouin zone (a truncated octahedron), but on a cube twice its size.}
	\label{fig:CMG}
\end{figure*}

As illustrated in Fig.~\ref{fig:CMG}, Co$_2$MnGa contains six linked NLs (red) which are shaped like flower petals (Fig.~\ref{fig:CMG}(a)). They touch smaller, circular nodal loops (blue), dubbed ``outer chains'', as shown in Fig.~\ref{fig:CMG}(b). These blue loops are, in turn, in contact with four smaller, also circular nodal loops (green), as in Fig.~\ref{fig:CMG}(c), forming another set of outer chains. Finally, the green rings touch the red flower petals, constituting an ``inner chain'' (Fig.~\ref{fig:CMG}(d)). All these features of Co$_2$MnGa contribute to a strongly non-linear response curve with large HHG coefficients, as presented in Figs.~4(d) and (e). These results were computed from a tight-binding model fitted to first-principle calculation results~\cite{CMG_theory}. 

The characteristically kink response curve of Co$_2$MnGa gives rise to much larger $n=3$ and $5$ harmonics than from YH$_3$ and Ti$_3$Al, and can be pedagogically explained via a detailed breakdown of its various topological NL linkages. We shall focus on the six red flower petals and the six blue rings, which are the dominant current contributors (largest features in Fig.~6(j)). Their individual contributions can be isolated through a 2-band approximate model which is amenable to analytic decomposition as follows:
\begin{equation}
H_\text{Co$_2$MnGa}= Re[f(\bold k)] \sigma_1 +Im[f(\bold k)]\sigma_3,
\end{equation}
where $f(\bold k)=\Pi_{n=1}^6 (u_n + \mathrm{i} v_n)$, which is a product of constituent nodal structures described by
\begin{align}
u_\text{even}\,&=&[\cos^3 (k_x/2)+\cos^3 (k_y/2) + \cos^3 (k_z/2) - m],\notag\\
u_\text{odd}\,&=&[\cos^3 (k_x/2)+\cos^3 (k_y/2) + \cos^3 (k_z/2) + m]
\end{align} 
and $v_{1,2}=\sin (k_x/2)$, $v_{3,4}=\sin (k_y/2)$, $v_{5,6}=\sin (k_z/2)$, the parameter $m=-0.5$ chosen so that the Fermi surface approximates that of \CMG. Remarkably, such a simple construction yields a rather accurate reconstruction of the Fermi surface of Co$_2$MnGa, as illustrated by its cross section comparison in Fig.~6(e) and the 3D superimposition in Fig.~5(a). 

Based on this approximate model, we can extract and compare the response contributions of the individual dominant NLs of CoMn$_2$Ga, which are the six flower petals and rings re-labeled cyan, magenta, blue, yellow, red and green in the top panel of Fig.~5(a). Interestingly, for an applied driving field along one of the rectangular coordinate axes i.e. $\hat y$, all these nodal structures obey only two possible response curves (type I: yellow and blue, type II: magenta, green, red, cyan). This follows from the cubic symmetry of \CMG, and enables a transparent reconstruction of the overall response. 
The two possible responses are presented in Figs.~5(b) and (c), where the green and yellow response curves correspond respectively to the contributions of the representative green and yellow nodal structures from Fig.~5(a). Curves from the other structures are equivalent to one of these, and are omitted for brevity. Together, both response curves exhibit large kinks, with their contrasting behavior at larger fields interfering to give rise to even more pronounced non-linearity and hence HHG properties, as detailed in the Methods. Their total contribution, plotted in Fig.~5(a), indeed qualitatively agrees with the full response curve from Fig.~4(d) despite considerable simplifications~\footnote{Not only were various smaller nodal features omitted from the dispersion, the Fermi regions were also restricted to the individual features.}. Its far larger non-linearity compared to that of YH$_3$ attests to the fact that to yield strong HHG, NLs must \emph{topologically link} and not merely intersect each other.


\noindent\textbf{Discussion --} In this work, we uncovered a very physically measurable consequence of topological nodal linkages, namely that they enhance and protect the HHG of optical radiation, which can be useful for Terahertz applications. As such, the sophisticated topology of nodal knots and link networks are no longer mere mathematical curiosities, but are in principle reconstructible from nonlinear response data in various directions. By venturing beyond the perturbative regime, we are able to access the effect of field impulses comparable to the size of NLs, and unveil the key role of nontrivial nodal linkages in enhancing the HHG already known to exist in simple nodal structures~\cite{Morimoto,Zyuzin}. 

Our results rigorously hold in the ballistic limit, which corresponds to the Terehertz regime for scattering times expected of high mobility samples of known nodal materials. Due to the robustness of our HHG mechanism, the characteristically superior HHG from nodal linkages is likely to extent beyond the ballistic and semi-classical limits, at least until hot phonon or interband scattering dominate. Despite our semi-classical analysis, our results crucially relies on the Fermi sea condensation of electrons, which is fundamentally due to the \emph{quantum mechanical} exclusion principle. As such, they are distinct from signatures of nodal geometry proposed in purely classical settings, such as topolectrical resonances due to nodal drumhead states~\cite{luo2018topological,lee2019imaging,li2019boundary}\footnote{However, topolectrical circuits that are intrinsically nonlinear~\cite{wang2019topologically} can also exhibit higher harmonic generation.}.

Reassuringly, the HHG enhancement by nodal linkages is also in consonance with quantitative response computations of three known representative nodal materials Ti$_3$Al, YH$_3$ and Co$_2$MnGa. In particular, the essential role of nodal linkages cannot be more evident from the comparison between Co$_2$MnGa and YH$_3$: Co$_2$MnGa, which contains multiple nodal linkages, exhibits far stronger HHG than YH$_3$, which contains multiple nodal intersections.


\begin{acknowledgments}
\noindent\textbf{Acknowledgements --} 
We thank Guoqing Chang, Yi Wei Ho, Ang Yee Sin, and Justin Song for discussions. J.G. is supported by the Singapore NRF Grant No. NRF-NRFI2017-04 (WBS No. R-144-000-378-281).  
\end{acknowledgments}

\bibliography{references_new,bibApril2019}

\appendix
\pagebreak
\begin{center}
\textbf{\large Methods and Appendices }
\end{center}

\setcounter{equation}{0}
\setcounter{figure}{0}
\setcounter{table}{0}
\setcounter{section}{0}
\makeatletter
\renewcommand{\theequation}{S\arabic{equation}}
\renewcommand{\thefigure}{S\arabic{figure}}
\renewcommand{\thesection}{S\Roman{section}}

\section{Validity of the ballistic approximation}

To justify that the ballistic regime $\Omega\tau\gg 1$ we assumed indeed corresponds to the Terahertz regime in reasonably clean samples of known nodal materials, we discuss below how the scattering time $\tau$ can be estimated. Since experimental data on $\tau$ is rare, especially for nodal materials which are relative novel, we shall
content ourselves with the following estimation approach. First, consider a representative single nodal loop (e.g. Ca$_3$P$_2$~\cite{Ca3P2}). For small loop radius, one can work with a continuum model and hence in toroidal coordinates. Consider Coulomb impurities as the dominant scatterers, and take the scatterer concentration to be $\sim 10^{24}\,$m$^{-3}$, i.e. of the same order as that of Weyl semimetals such as Cd$_3$As$_2$~\cite{Shaffique}. For screening, we take the long-wavelength limit to obtain a Thomas-Fermi dielectric function $\varepsilon(\bm{q})=1+\frac{q_s^2}{\bm{q}^2}$, where $q_s^2=e^2\nu(E_F)/\varepsilon_0$ is the square of the inverse screening length, with $\nu$ the density of states, $\varepsilon_0$ the vacuum permittivity, and $e$ the elementary charge. To obtain a rough estimate for the relaxation time, one can proceed by assuming that the relaxation time depends only on the Fermi energy. Denoting it as $\tau_{\bm{k}}$, it can can then be calculated~\cite{AnisotropicRelaxationTime} numerically by solving
 \begin{equation}
 \begin{split}
 1 &= \int \frac{\mathrm{d}^3\bm{k}'}{(2\pi)^3} \frac{2\pi}{\hbar}\vert\langle \bm{k}'|\hat{V}|\bm{k}\rangle \vert^2 \delta(\epsilon_{\bm{k}}-\epsilon_{\bm{k}'}) \left(\tau_{\bm{k}}^{(i)}-\frac{v_{\bm{k}'}^{(i)}}{v_{\bm{k}}^{(i)}}\tau_{\bm{k}'}^{(i)}		\right).
 \end{split}
 \end{equation}
 for anisotropic band dispersion $\epsilon_{\bm{k}}$. In the above, $\hat{V}$ is the screened impurity potential, and the superscript $(i)$ indicates the direction of the applied field. The resulting relaxation time depends on the Fermi energy, but its values can be placed within the range $\tau\,\sim\, 5\times 10^{-13}\,s$, which is similar to that of Graphene as obtained via RF admittance measurements~\cite{TransportTimeGraphene}. Then, with a terahertz driving field of frequency $\sim 100\,$THz, one has, roughly, $\Omega\tau\,\sim\, 50$, validating the ballistic approximation.



\section{Proof of the non-monotonicity of the response of a topological linkage}
To show that singularities of the $\bold v=\nabla_{\bold k}\epsilon$ vector field can indeed ensure significant non-linearity in the current response $\bold J(\bold A)$, we approximate the Fermi-Dirac distribution $F(\epsilon)$ as a step function\footnote{This introduces negligible errors at room temperature.}, and rewrite the velocity current contribution in a chosen $\hat e$ direction as 
\begin{eqnarray}
 J_{\hat e}(\bold A)&=&\int F(\epsilon(\bold k-\bold A))\frac{\partial \epsilon(\bold k)}{\partial k_\parallel} d^3\bold k \notag\\
&\approx&\int \theta(\mu-\epsilon(\bold k)) \left(\frac{\partial \epsilon(\bold k+\bold A)}{\partial k_\parallel} d k_\parallel\right ) d\vec k_\perp\notag\\
&=& \int\sum_{\alpha} \left[\epsilon\left(\bold k^\alpha+\Delta \bold k_\parallel^\alpha+\bold A\right)-\epsilon\left(\bold k^\alpha-\Delta \bold k_\parallel^\alpha+\bold A\right)\right] d\vec k_\perp\notag\\
&=& \int\sum_{\alpha\in \text{nodes}} \Delta \epsilon_\alpha(\vec k_\perp^\alpha+\bold A) d\vec k_\perp
\end{eqnarray}
where $k_\parallel=\bold k_\parallel\cdot \hat e = \bold k\cdot \hat e$ and $\vec k_\perp=\bold k-\bold k_\parallel$ are the components of $\vec k$ parallel and perpendicular to $\hat e$. The BZ is sliced into strips parallel to $\hat e$ and indexed by $\bold k_\perp$, such that the thin nodal Fermi tube is mathematically broken into thin cross sections $\alpha$ in each $\bold k_\perp$ strip. Each $\alpha$ has a width of $2|\Delta \bold k_\parallel^\alpha|$ that is determined via $\epsilon(\bold k^\alpha\pm \Delta \bold k_\parallel^\alpha)=\mu$, where $\bold k^\alpha$ is the central position of $\alpha$. The current is thus proportional to the sum of all energy differences $\Delta \epsilon_\alpha(\vec k_\perp^\alpha) =\epsilon\left(\bold k^\alpha+\Delta \bold k_\parallel^\alpha+\bold A\right)-\epsilon\left(\bold k^\alpha-\Delta \bold k_\parallel^\alpha+\bold A\right)$ between the two ``sides'' of the Fermi tube. This rewriting of current response as shifted boundary terms actually remains valid in arbitrarily high dimensions~\cite{lee2018electromagnetic}, and can be elegantly expressed in terms of generalized Bianchi identities~\cite{petrides2018six}.

Nonlinear response in $J_{\hat e}$ can be investigated through the gradient $\frac{d J_{\hat e}}{d  A_{\hat e}}=$
\begin{eqnarray}&&\sum_{\alpha}\frac{d}{d  k^\alpha_\parallel}\int \left[\epsilon\left(\bold k^\alpha+\Delta \bold k_\parallel^\alpha+\bold A\right)-\epsilon\left(\bold k^\alpha-\Delta \bold k_\parallel^\alpha+\bold A\right)\right] d\vec k_\perp\notag\\
&\approx &2 \int\sum_{\alpha}\Delta k^\alpha_\parallel\frac{d^2 \epsilon(\vec k^\alpha_\perp+\bold A)}{d k_\parallel^2} d\vec k_\perp\notag\\
&= &2 \int\sum_{\alpha}\frac{\mu}{\hat e\cdot\bold v_F}\frac{d^2 \epsilon(\vec k^\alpha_\perp+\bold A)}{d k_\parallel^2} d\vec k_\perp
\label{dJdp0}
\end{eqnarray}
where $\hat e\cdot \bold v_F$ is the component of the Fermi velocity of NL $\alpha$ along the applied field. More generically, the response tensor can be obtained by replacing the second derivative by the Hessian, and projecting the Fermi velocity along the applied field. To summarize, Eq.~\ref{dJdp0} expresses the current response gradient as the sum of the dispersion curvature over all occupied states, inversely weighted by the Fermi velocity.
\begin{figure}
\includegraphics[width=\linewidth]{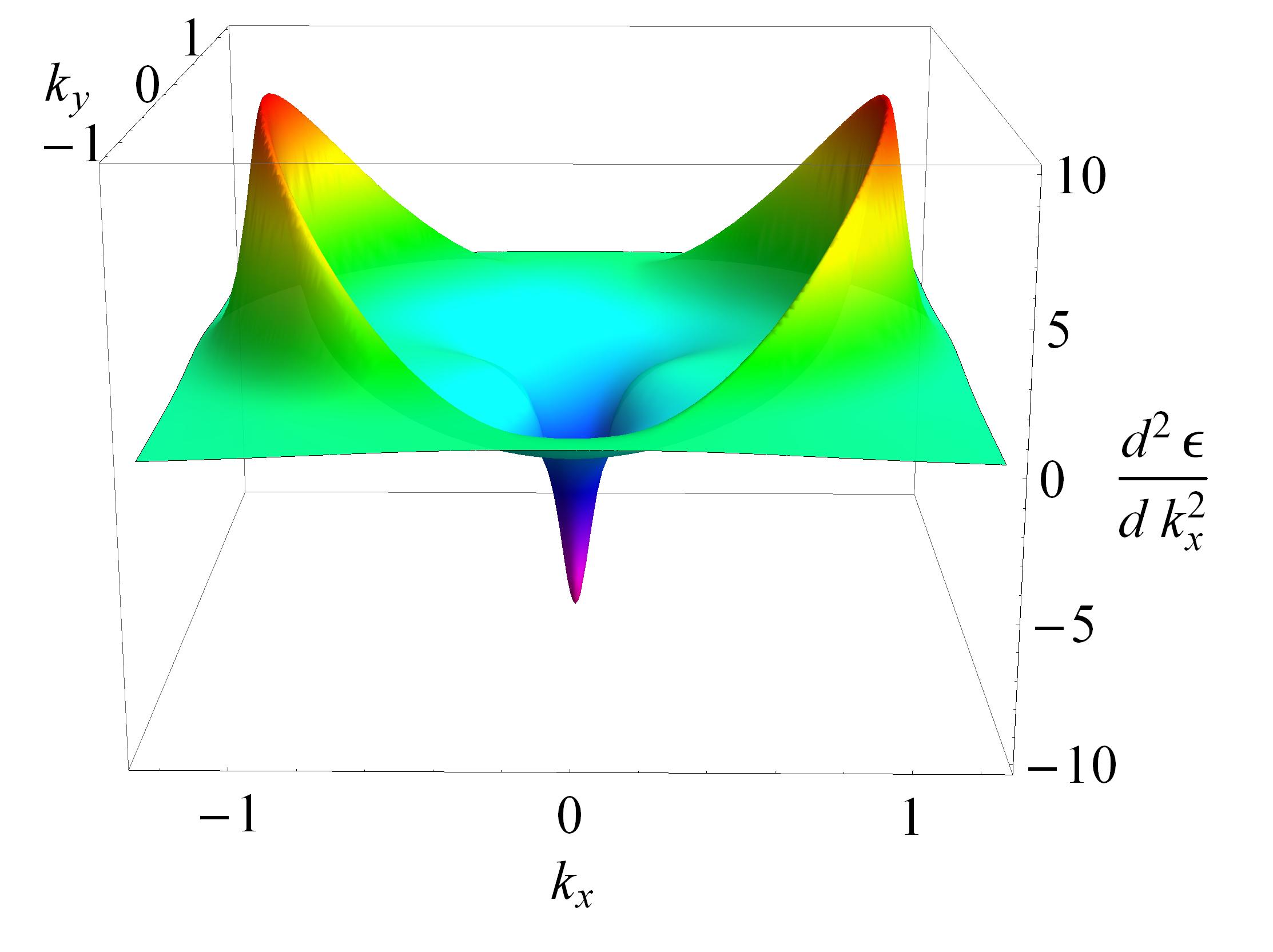}
\caption{(Color online) Plot of the curvature $\frac{d^2 \epsilon}{dk_x^2}$ for the Hopf NL system (Eq. \ref{diracE}) in the plane of one of its loops, where $k_z=0$. There are peaks of height $\sim \frac1{m_r}$ around the ``ears'' surrounding the (unit) nodal circumference of the loop, and an sharp trench of depth $\sim\frac1{m_0}$ arising from the other perpendicularly impinging loop.}
\label{fig:bandstructure}
\end{figure}

\section{Alternative model for the single nodal loop}

\begin{figure}[H]
\begin{minipage}{\linewidth}
\subfloat[]{\includegraphics[width=0.5\linewidth]{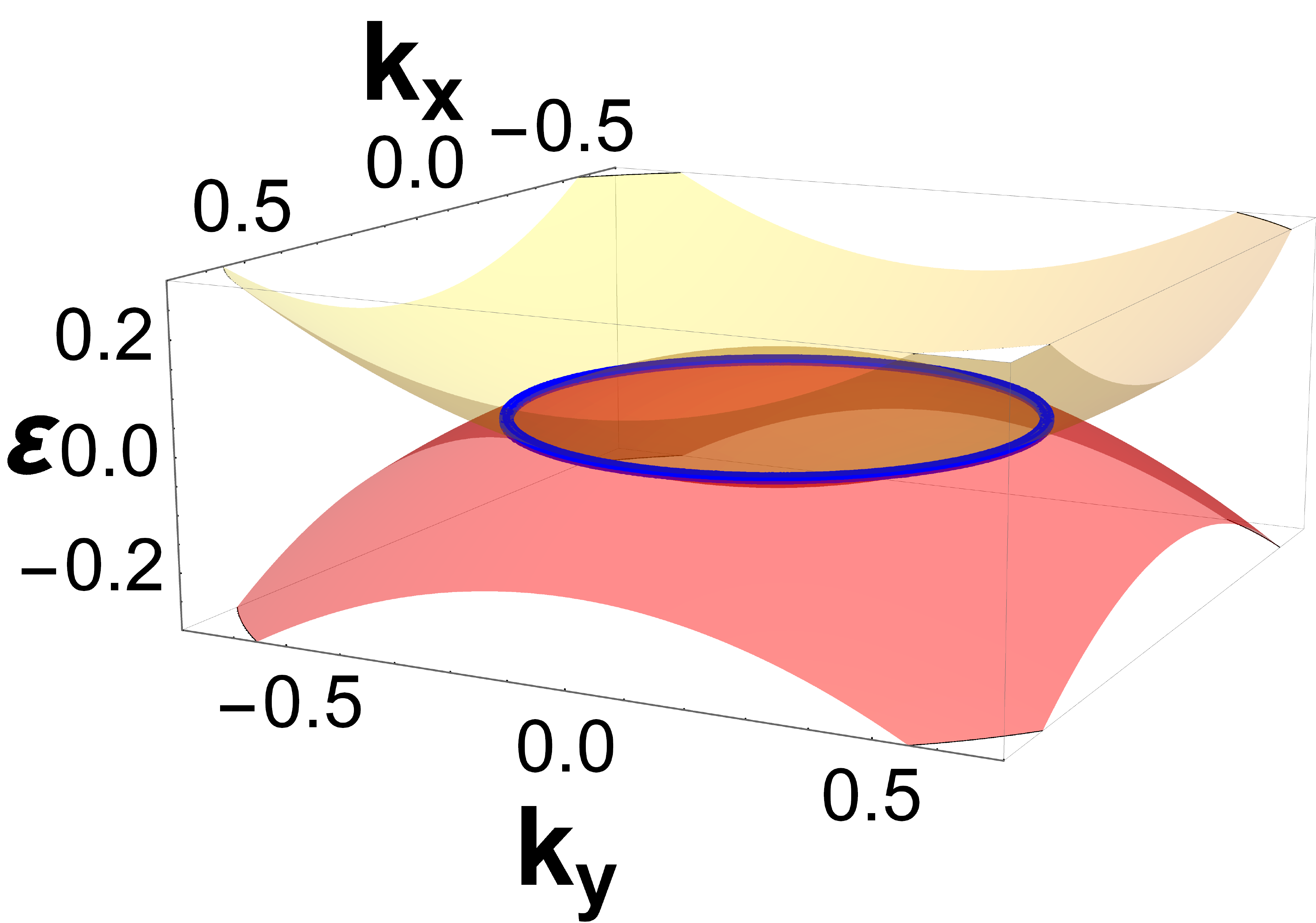}}
\subfloat[]{\includegraphics[width=0.5\linewidth]{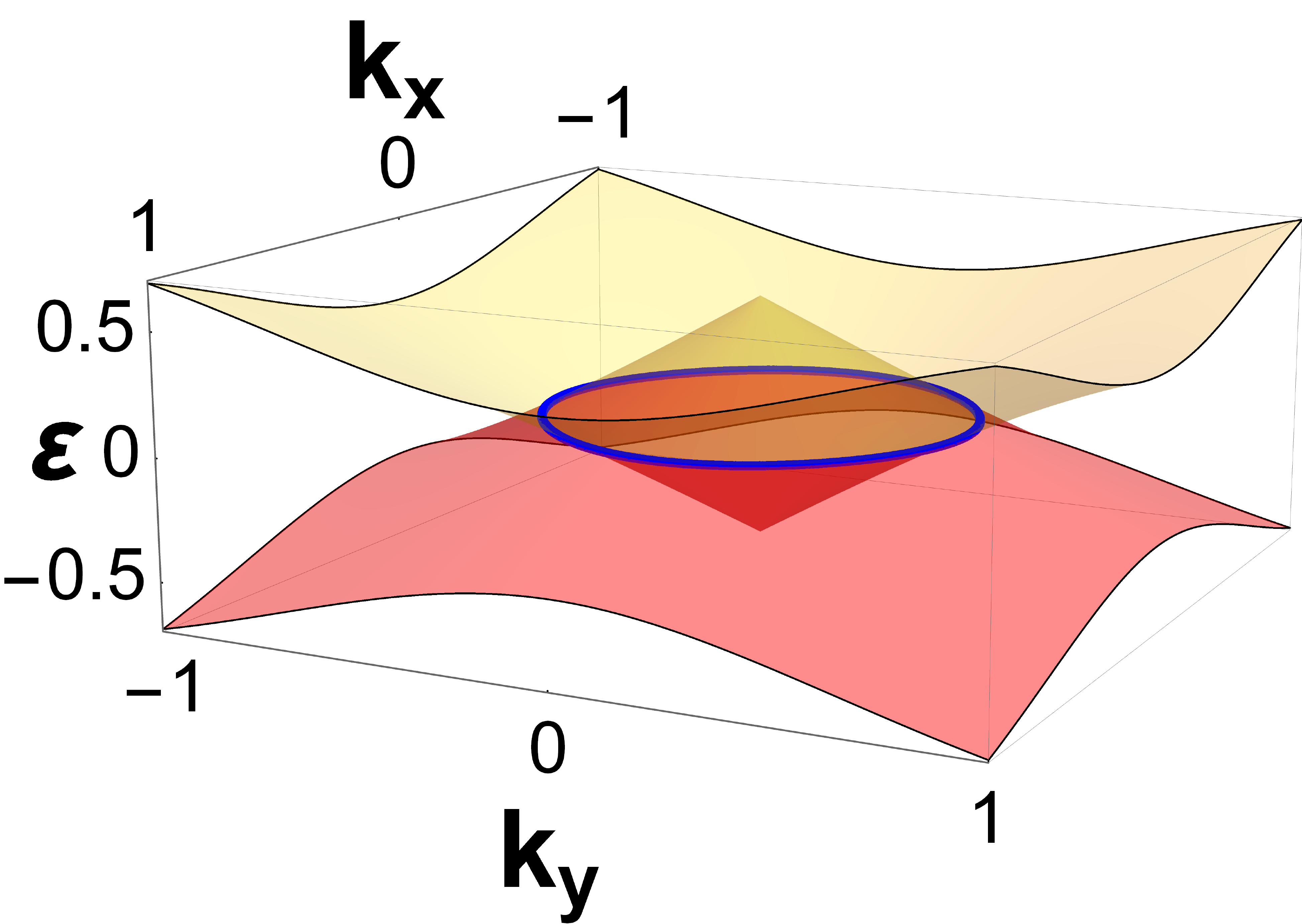}}
\end{minipage}
\caption{Conduction (blue) and valence (yellow) bands intersect at a single NL (red) in the (a) 2-band single NL model from the main text, and (b) the $\epsilon_{1,2}$ subspace of of the 4-band single NL model introduced above. The 4-band model contains a divergent curvature at the center of the ring, which gives rise to enhanced though unprotected HHG.
}
\label{fig:24bands}
\end{figure}

In the main text, we have mentioned that a single NL has a response nonlinearity that is ``unprotected''. This is because its dispersion $\epsilon(\bold k)$ can be deformed, without affecting its ring of nodes, such that its velocity vector field $\bold v=\nabla_{\bold k}\epsilon$ exhibits very different cancellation extents. To illustrate this, we introduce an alternative toy model for the single NL:

\begin{eqnarray}\label{8}
H_\text{alt}&=&\sin{k_x}\sigma_x\otimes1+\sin{k_y}\sigma_y\otimes\sigma_y+\sin{k_z}\sigma_z\otimes1+b\sigma_x\otimes\sigma_x\nonumber\\&=&\begin{bmatrix}\sin{k_z}&0&\sin{k_x}&b-\sin{k_y}\\0&\sin{k_z}&b+\sin{k_y}&\sin{k_x}\\\sin{k_x}&b+\sin{k_y}&-\sin{k_z}&0\\b-\sin{k_y}&\sin{k_x}&0&-\sin{k_z}\\\end{bmatrix}
\end{eqnarray}
where $0<b<1$. This 4-band model possesses energy bands 
\begin{equation}\label{e1_2}
\epsilon_{1,2}=\pm\sqrt{\sin^2{k_z}+(\sqrt{\sin^2{k_x}+\sin^2{k_y}}-b)^2},
\end{equation}
\begin{equation}\label{e3_4}
\epsilon_{3,4}=\pm\sqrt{\sin^2{k_z}+(\sqrt{\sin^2{k_x}+\sin^2{k_y}}+b)^2}
\end{equation}
The intersection of the bands $\epsilon_{1,2}$ of Eq.~\ref{e1_2} forms a single NL in the $k_x$-$k_y$ plane. But due do their proximity with bands $\epsilon_{3,4}$, band intersections also exist at the origin, imitating the effect of another node at the center of the NL in the $k_x$-$k_y$ plane despite the absence of another linked NL. A comparison with the dispersion of the 2-band single NL model is shown in Fig.~\ref{fig:24bands}. That said, the $\bold v$ singularity at the center can be easily deformed away from the plane of the NL, thereby reducing the non-linear response, unlike in the Hopf NL where the central singularity is an unavoidable feature of the nodal linkage.
\section{Further details on the nodal materials and their lattice models}

\subsection{Ti$_3$Al}

The energy dispersion of Ti$_3$Al is modeled by the expression~\cite{zhang2018nodal}
\begin{equation}\label{Ti3Al}
\epsilon_\text{Ti$_3$Al}(\bold k)=\frac{1}{2}\bigg(h_{1}+h_{2}+\sqrt{(h_1-h_2)^2+h^2}\bigg)
\end{equation}
where $h_{i}=A_i(k_x^2+k_y^2)+B_ik_z^2+M_i$ and $h=2Ck_z$ with parameter values $A_1=-9.66\text{eV}$\AA$^{2}$, $A_2=11.37\text{eV}$\AA$^{2}$, $B_1=36.22\text{eV}$\AA$^{2}$, $B_2=-25.71\text{eV}$\AA$^{2}$, $M_1=0.12\text{eV}$, $M_2=-0.52\text{eV}$ and $C=22.34\text{eV}$\AA$^{2}$.

To conveniently represent its cross section in the $k_z=0$ plane, it is also useful to use its piecewise polynomial approximation
\begin{align} 
\epsilon_\text{Ti$_3$Al}|_{\sqrt{k}\leq 0.16}\,&\approx \,c_1+c_2k\notag\\
\epsilon_\text{Ti$_3$Al}|_{\sqrt{k}>0.16}\,&\approx\, c_3+c_4k+c_5k^2+c_6k^3+c_7k^4,
\end{align}
where $k=k_x^2+k_y^2$ and $c_1=0.0911$, $c_2=- 4.0615$, $c_3=-0.2339$, $c_4=9.715$, $c_5=-41.8543$ $c_6=81.1788$ and $c_7= - 59.1387$. Units of energy and momentum $k$ are in $eV$ and $\AA^{-1}$ respectively.
\begin{figure}[H]
\begin{minipage}{\linewidth}
\subfloat[]{\includegraphics[width=0.52\linewidth]{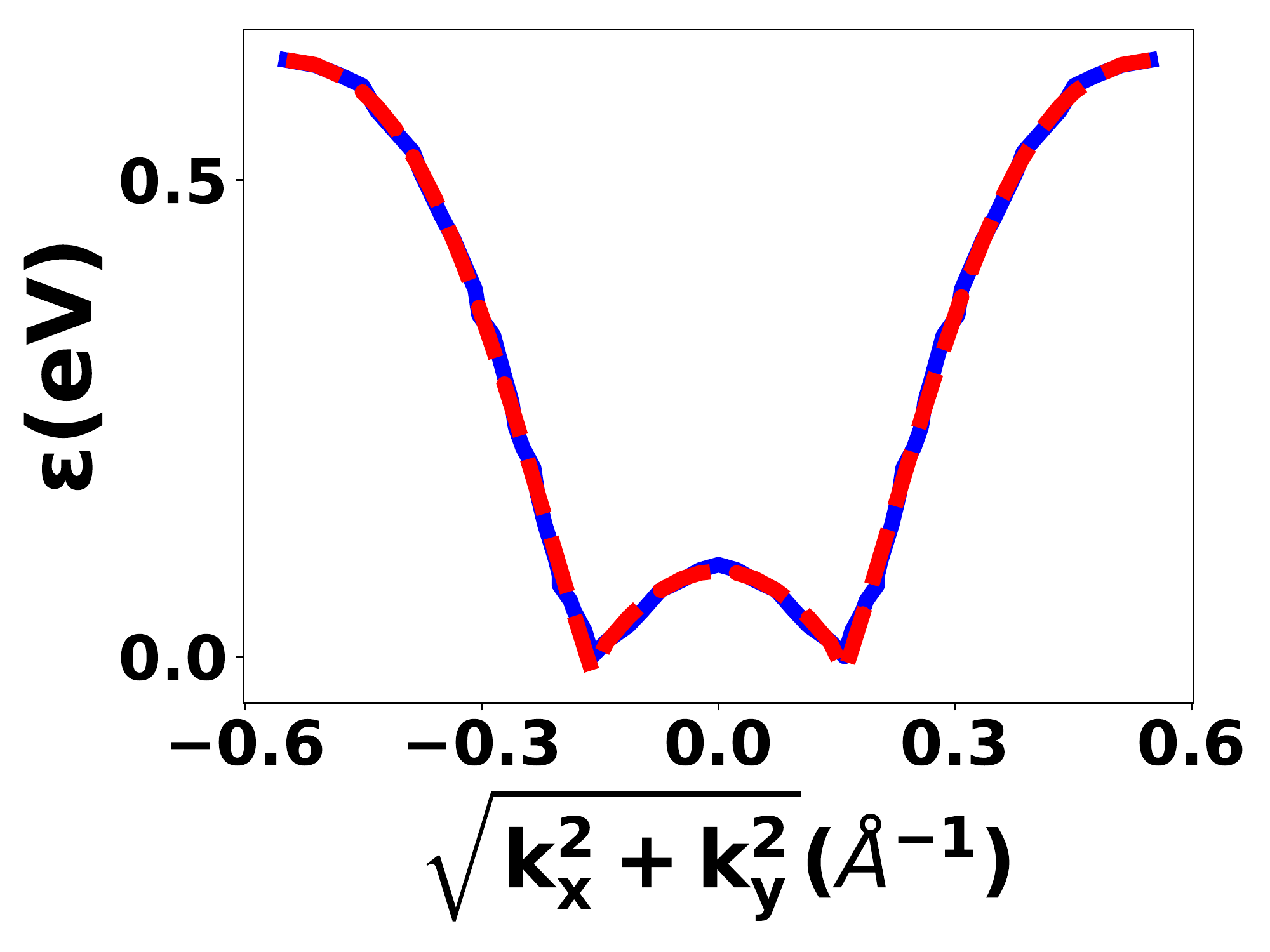}}
\subfloat[]{\includegraphics[width=0.48\linewidth]{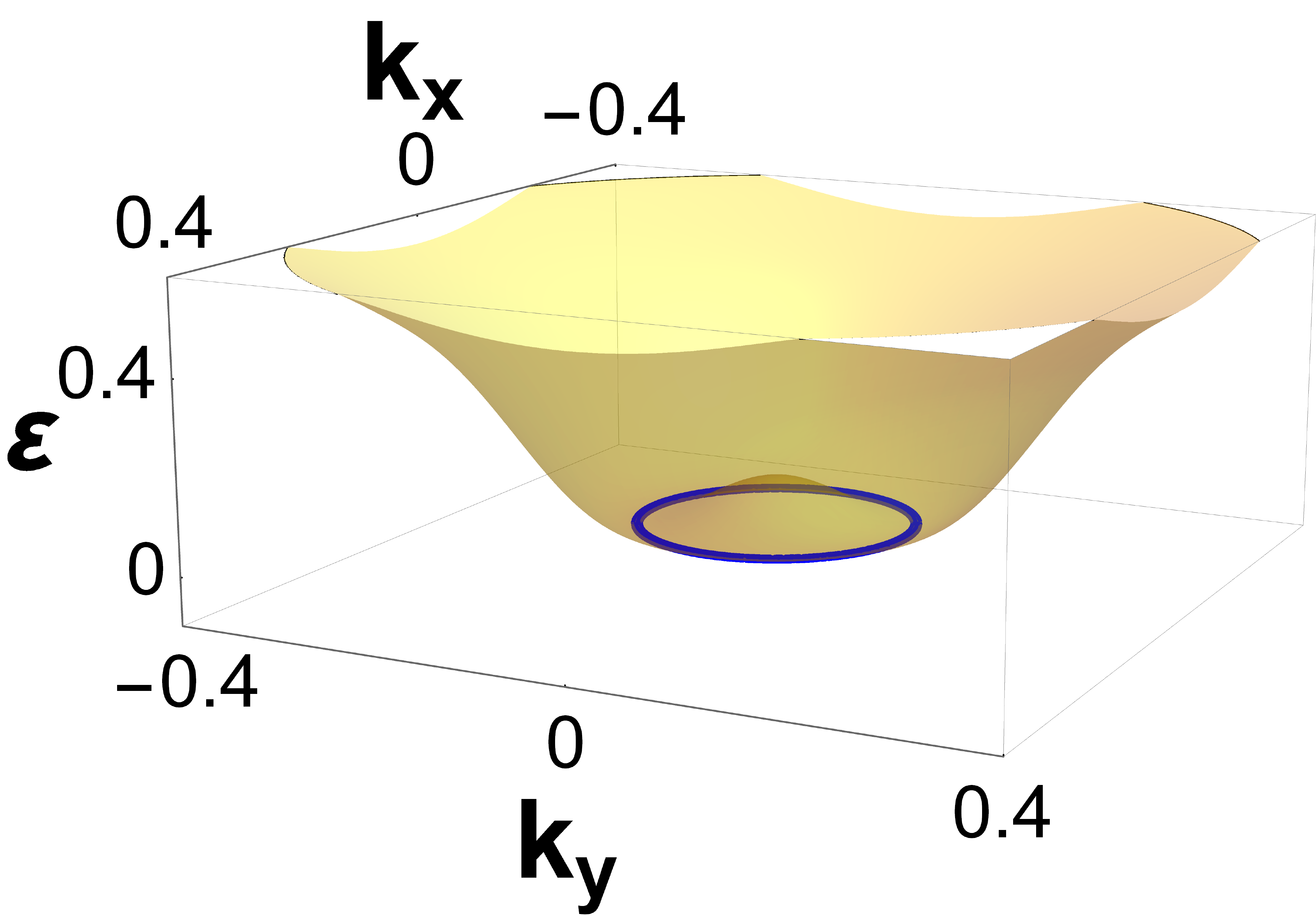}}
\end{minipage}
\caption{(a) Comparison of actual dispersion data of Ti$_3$Al (blue and the best fit piecewise curve (red dashed) along the radial direction in momementum space, i.e. $k=\sqrt{k_x^2+k_y^2}$; (b) A distinct NL lies at the intersection of the $\pm\,\epsilon_\text{Ti$_3$Al}(k_x,k_y,0)$ bands, i.e at $\epsilon(k_x,k_y,0)=0$.
}
\end{figure}

\subsection{YH$_3$}
\begin{figure}
\begin{minipage}{\linewidth}
\subfloat[]{\includegraphics[width=0.5\linewidth]{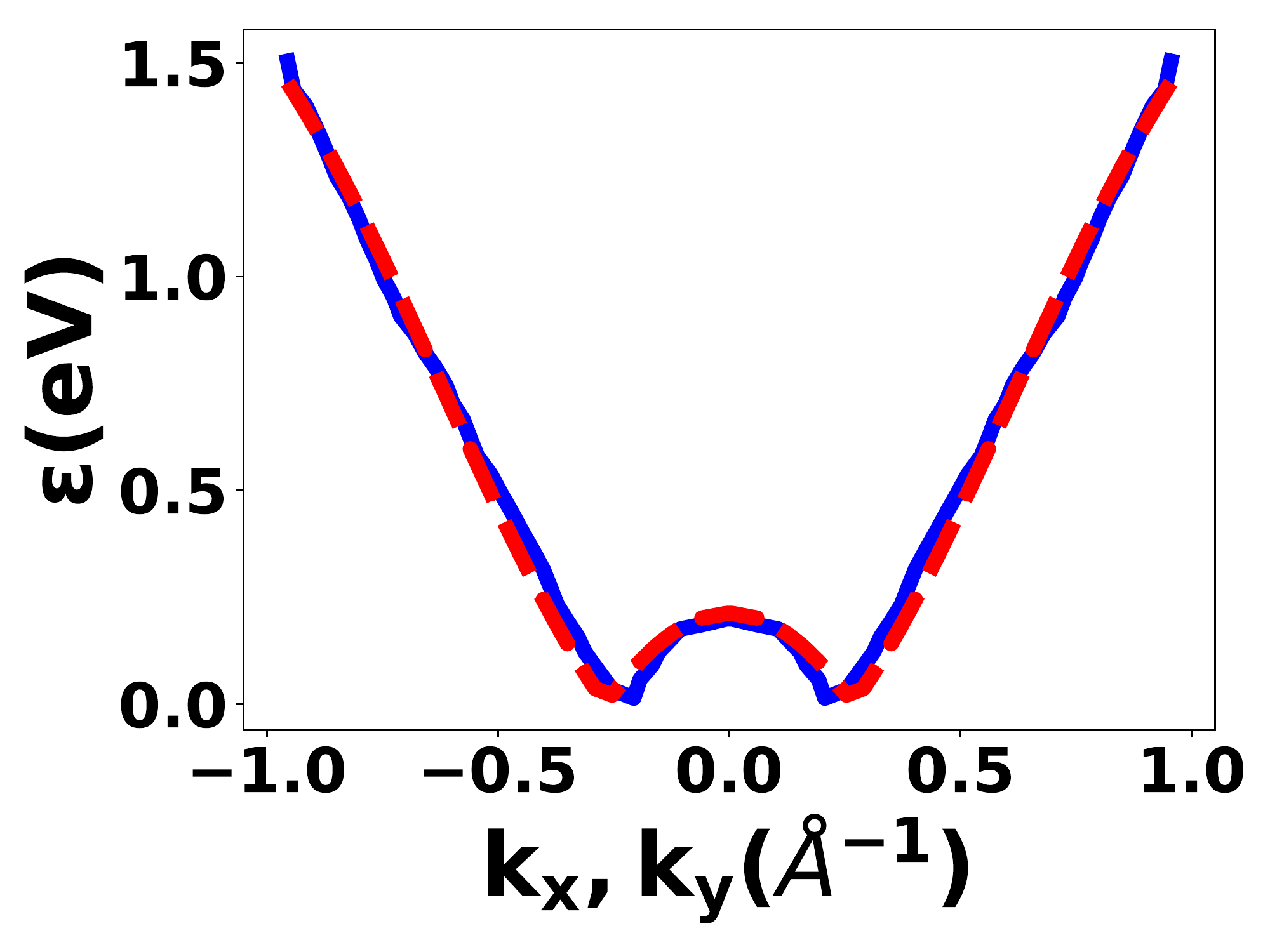}}
\subfloat[]{\includegraphics[width=0.5\linewidth]{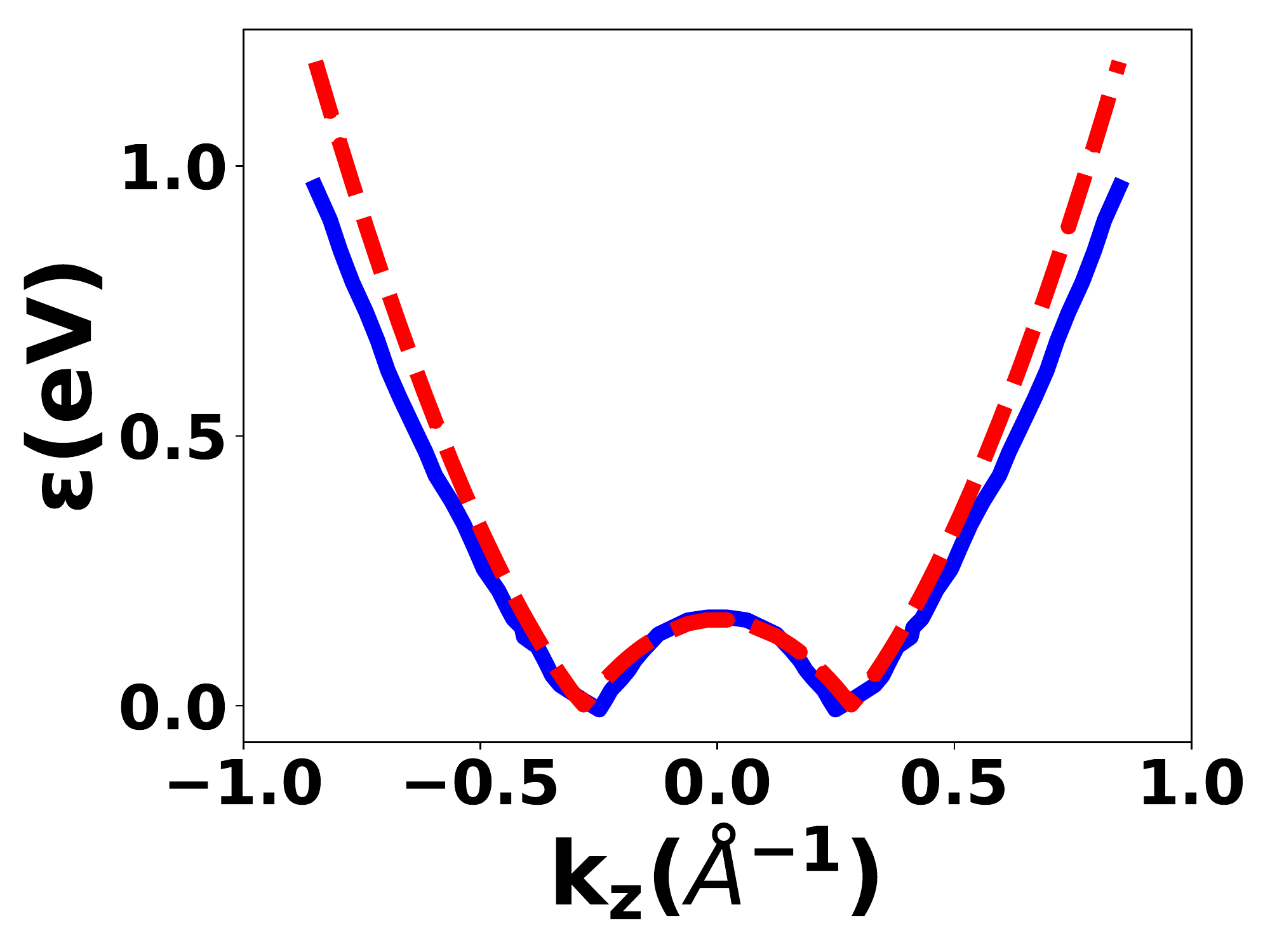}}\\
\subfloat[]{\includegraphics[width=0.5\linewidth]{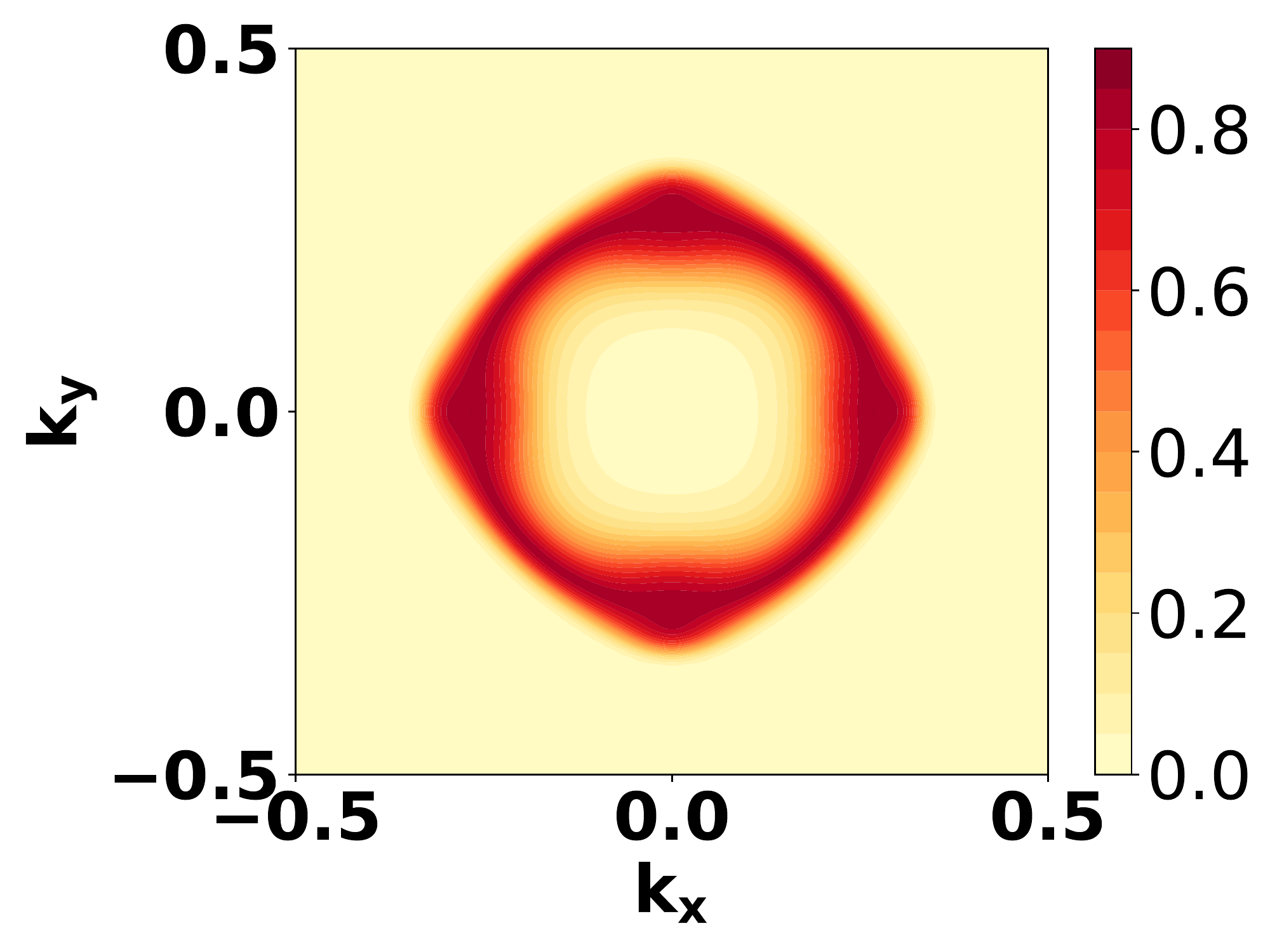}}
\subfloat[]{\includegraphics[width=0.5\linewidth]{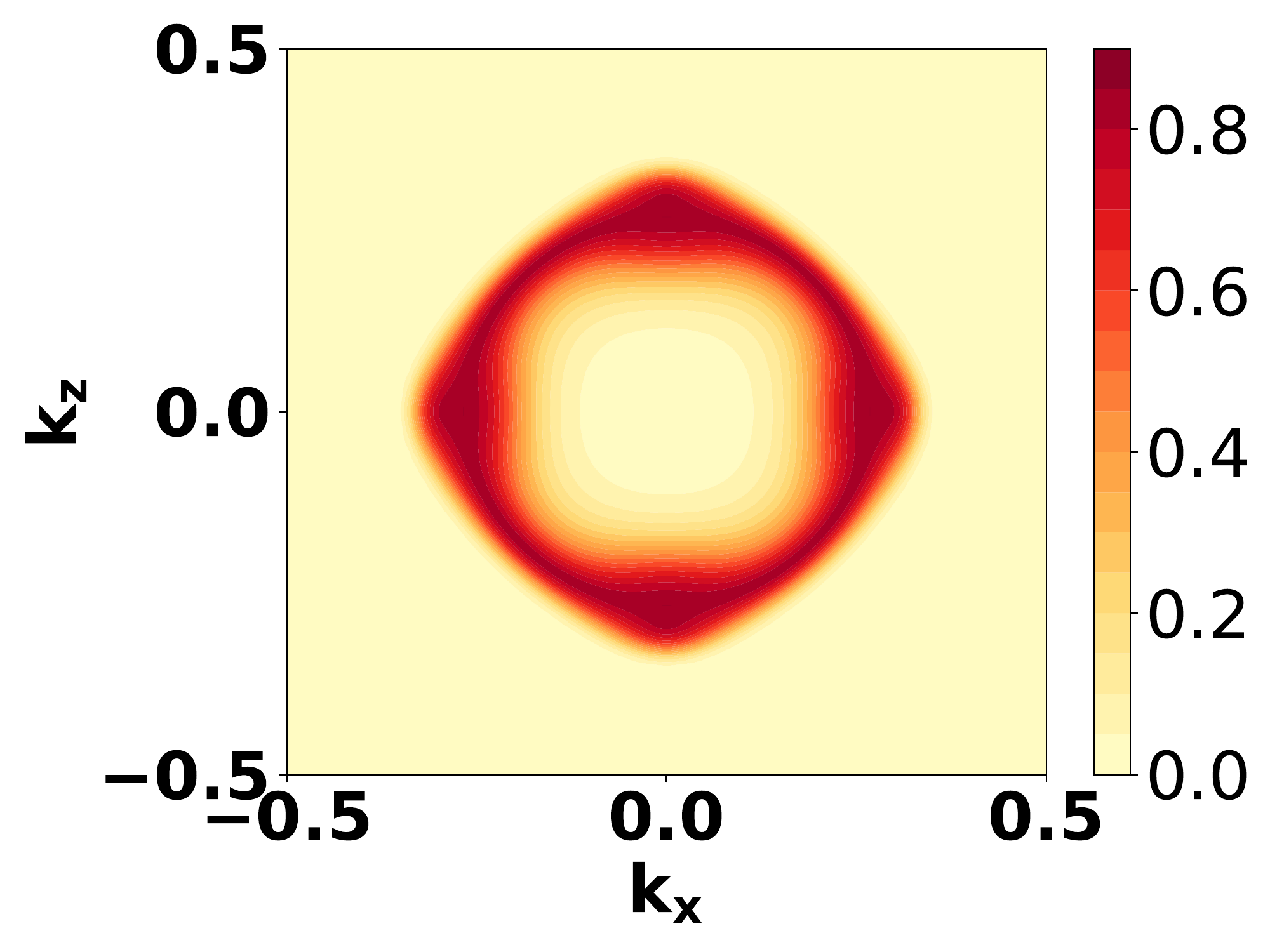}}
\end{minipage}
\caption{(a-b) Comparison of actual dispersion data of YH$_3$ (blue) and the best fit curve (red dashed) along (a) either the $k_x$ or $k_y$ direction, and (b) the $k_z$ direction; (c-d) Fermi regions of the YH$_3$ with $\mu=0.015eV$ and $T=10$K along (c) the $k_x$-$k_y$ plane and (d) the $k_x$-$k_z$ plane. The color bar indicates the Fermi occupancy factor. 
}
\label{YH3app}
\end{figure}

YH$_3$ consists of three intersecting NLs along each orthogonal plane in the 3D BZ, and can be modeled by the dispersion
\begin{equation}
\epsilon_\text{YH$_3$}(\bold k)=\sqrt{(g_1^2+h_1^2)(g_2^2+h_2^2)(g_3^2+h_3^2)}
\end{equation}
where the factors $g_i^2+h_i^2$ each yield an individual NL~\cite{zhou2018hopf}. They are given by
$$g_1=\sin(k_z),g_2=\sin(k_x), g_3=\sin(k_y)$$
$$h_1=a_1(r_1(\cos k_x))^{3}+s_1(\cos k_y))^{3}+t_1(\cos z))^{3}-m_1)$$
$$h_2=a_2(\cos k_x+\cos k_y+\cos k_z-m_2)$$
$$h_3=a_3(\cos k_x+\cos k_y+\cos k_z-m_3)$$
where DFT-fitted parameters~\cite{shao2018nonsymmorphic} are given by $m_1=2.99$, $a_1=2$, $r_1=s_1=t_1=1.032$, $n_1=3$, $m_{2,3}=2.96$ and $a_{2,3}=4$. Units of energy and momentum $k$ are $eV$ and \AA$^{-1}$ respectively. This ansatz is sufficient in taking into account of the slight anisotropy in the NLs (Fig.~\ref{YH3app}).
 
	\subsection{Co$_2$MnGa}
	\subsubsection{Tight-binding model}

We take the tight-binding model of Co$_2$MnGa from Ref.~\cite{CMG_theory}, which we copy here for ease of reading. It consists of six bands: Three $d$-orbitals from Mn and three $p$-orbitals from Ga. In reciprocal space with the basis $(d_{xz},d_{yz},d_{xy},p_x,p_y,p_z)$, it is given by:
	\begin{equation}
	H(\bm{k})=\begin{pmatrix}
	\xi^d_1 & 0 & 0 & \xi^{dp}_{11} & 0 & \xi^{dp}_{13} \\
	0 & \xi^d_2 & 0 & 0 & \xi^{dp}_{22} & \xi^{dp}_{23} \\
	0 & 0 & \xi^d_3 & \xi^{dp}_{31} & \xi^{dp}_{32} & 0 \\
	\xi^{dp}_{11} & 0 & \xi^{dp}_{31} & \xi^p_1 & \xi^p_{12} & \xi^p_{31} \\
	0 & \xi^{dp}_{22} & \xi^{dp}_{32} & \xi^p_{12} & \xi^p_2 & \xi^p_{23} \\
	\xi^{dp}_{13} & \xi^{dp}_{23} & 0 & \xi^p_{31} & \xi^p_{23} & \xi^p_3
	\end{pmatrix}(\bm{k}),
	\end{equation}
	with:
	\begin{widetext}
	\begin{subequations}
	\begin{align}
	\xi^d_1(\bm{k}) &= 4t_1 \cos\frac{k_x}{2}\cos\frac{k_z}{2} + 2t_2(\cos k_x+\cos k_z) + 2t_3 \cos k_y + \epsilon_d \\
	\xi^d_{2}(\bm{k}) &=  4t_1 \cos\frac{k_y}{2}\cos\frac{k_z}{2} + 2t_2(\cos k_y+\cos k_z) + 2t_3 \cos k_x + \epsilon_d \\
	\xi^d_{3}(\bm{k}) &=  4t_1 \cos\frac{k_x}{2}\cos\frac{k_y}{2} + 2t_2(\cos k_x+\cos k_y) + 2t_3 \cos k_z + \epsilon_d \\
	\xi^p_{1}(\bm{k}) &=  4t_4 \cos\frac{k_y}{2}\cos\frac{k_z}{2} + 2t_5(\cos k_y+\cos k_z) + 2t_6 \cos k_x + \epsilon_p \\
	\xi^p_{2}(\bm{k}) &=  4t_4 \cos\frac{k_x}{2}\cos\frac{k_z}{2} + 2t_5(\cos k_x+\cos k_z) + 2t_6 \cos k_y + \epsilon_p \\
	\xi^p_{3}(\bm{k}) &=  4t_4 \cos\frac{k_x}{2}\cos\frac{k_y}{2} + 2t_5(\cos k_x+\cos k_y) + 2t_6 \cos k_z + \epsilon_p \\
	\xi^p_{12}(\bm{k}) &= -4t_7 \sin\frac{k_x}{2}\sin\frac{k_y}{2} \\
	\xi^p_{23}(\bm{k}) &= -4t_7 \sin\frac{k_y}{2}\sin\frac{k_z}{2} \\
	\xi^p_{31}(\bm{k}) &= -4t_7 \sin\frac{k_x}{2}\sin\frac{k_z}{2} \\
	\xi^{dp}_{11}(\bm{k}) &= \xi^{dp}_{22}(\bm{k}) = 2t_8 \sin\frac{k_z}{2}\\
	\xi^{dp}_{13}(\bm{k}) &= \xi^{dp}_{32}(\bm{k}) = 2t_8 \sin\frac{k_x}{2}\\
	\xi^{dp}_{23}(\bm{k}) &= \xi^{dp}_{31}(\bm{k}) = 2t_8 \sin\frac{k_y}{2}
	\end{align}
	where the fitte tight-binding parameters are given (in units of eV) by: $t_1=-0.31,t_2=-0.018,t_3=-0.01,t_4=0.2,t_5=-0.02,t_6=0.04,t_7=0.28,t_8=-0.34,\epsilon_d=-0.6,\epsilon_p=0.6$.
		\end{subequations}
\end{widetext}

\subsubsection{Velocity field in the nodal Co$_2$MnGa network}

\begin{figure}
	\begin{minipage}{\linewidth}
		\centering
		\subfloat[]{\includegraphics[width=.5\linewidth]{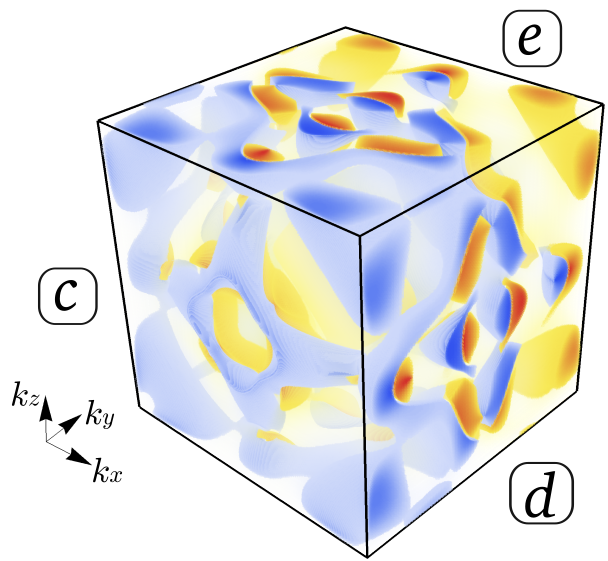}}
		\subfloat[]{\includegraphics[width=.5\linewidth]{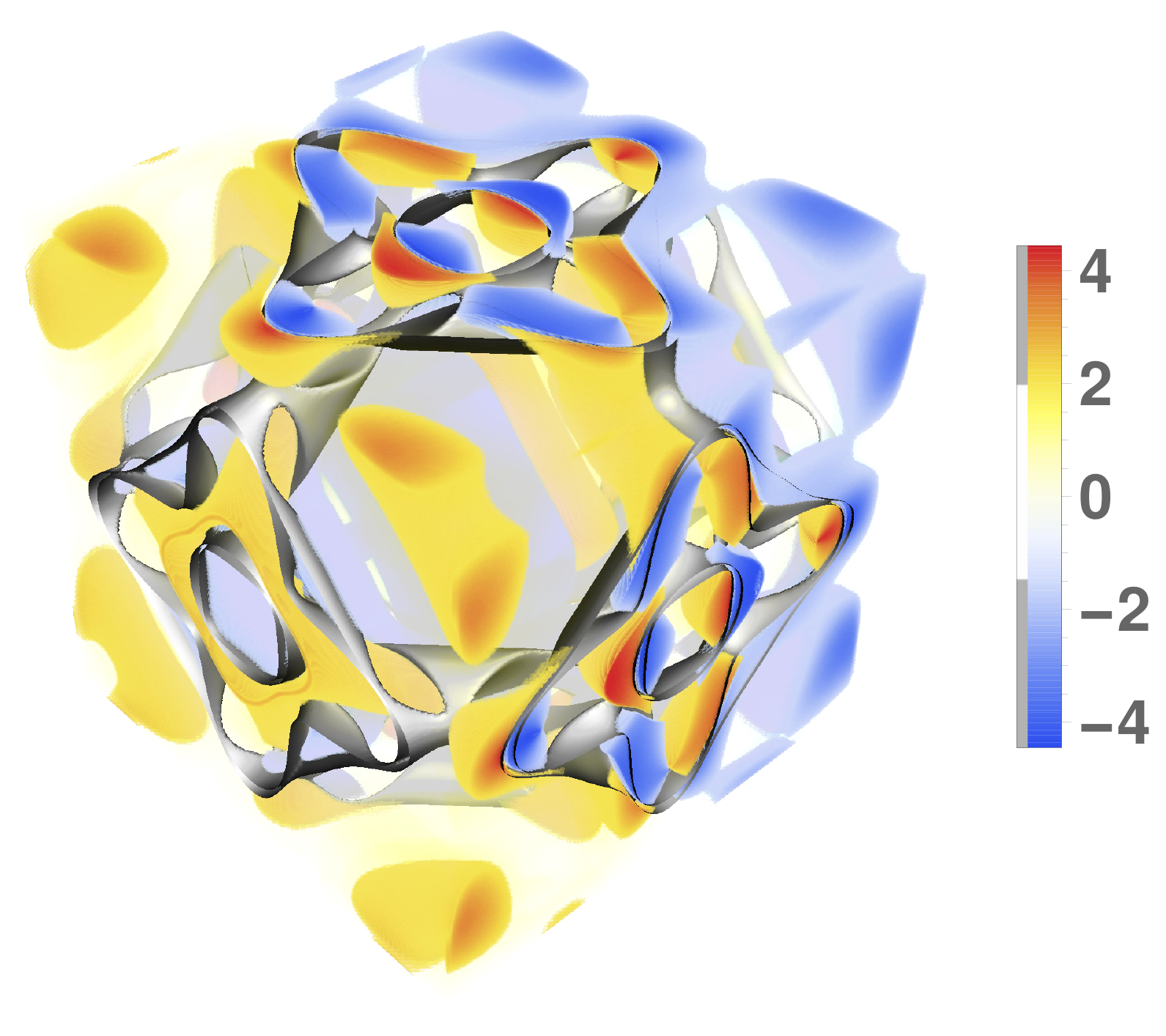}} \\
		\subfloat[]{\includegraphics[width=.33\linewidth]{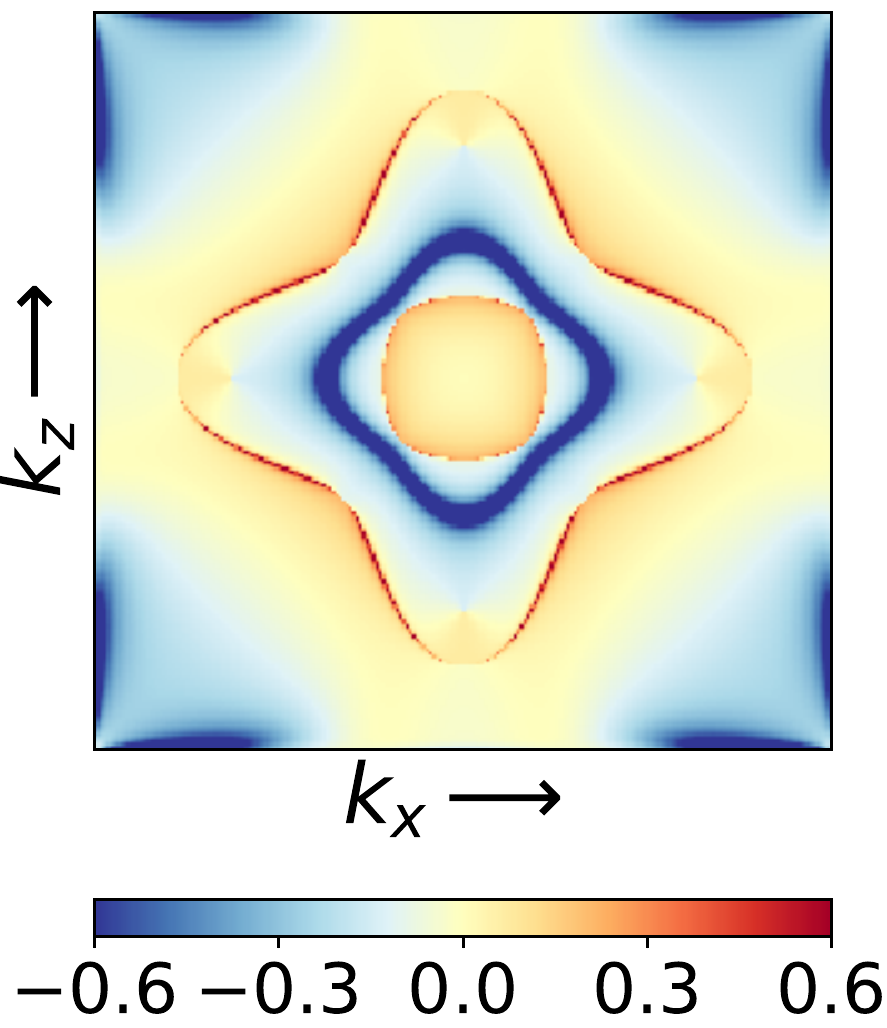}}
		\subfloat[]{\includegraphics[width=.33\linewidth]{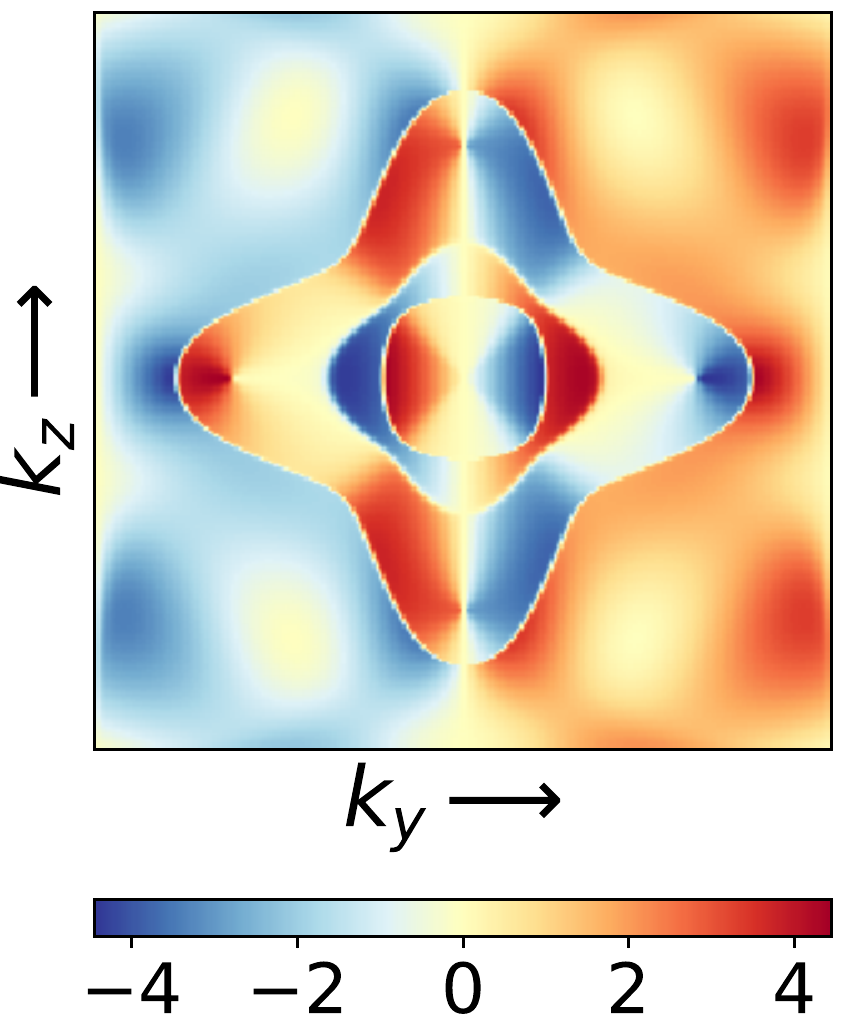}}
		\subfloat[]{\includegraphics[width=.33\linewidth]{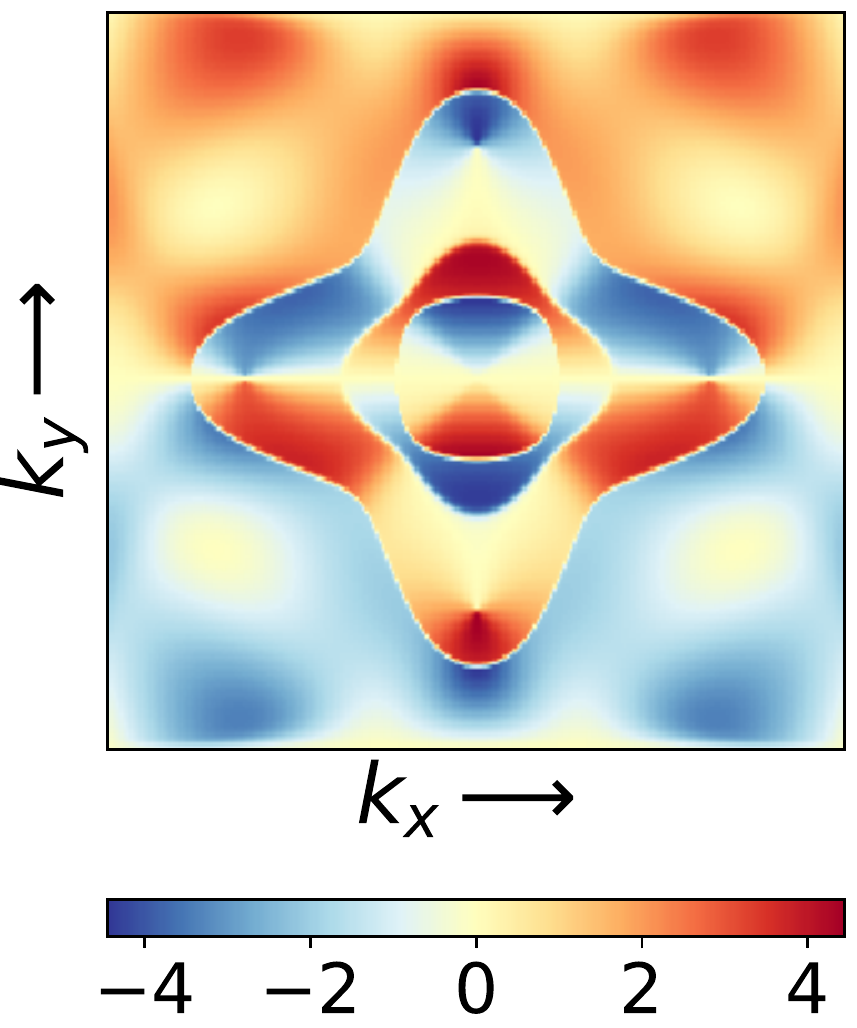}} \\
	\end{minipage}
	\caption{(a) The density plot of $v_y$, i.e. the $y$-component of the group velocity of the third band (ordered from high to low energies) of the tight-binding model of Co$_2$MnGa. The choice of the velocity component is arbitrary owing to the cubic symmetry of Co$_2$MnGa. The planes labeled (c-e) correspond to the cross sections on which we display $v_y$ below. (b) A rotated view of (a), overlaid by the Fermi surface of \CMG. (c) Density plot of $v_y$ restricted on the $k_z$-$k_x$ plane at $k_y=-2\pi/a$. (d) Density plot of $v_y$ on the $k_z$-$k_y$ plane at $k_x=2\pi/a$. (e) Density plot of $v_y$ on the $k_y$-$k_x$ plane at $k_z=2\pi/a$. All color bars are expressed in units of $\times 10^{5}\,$ms$^{-1}$.}
	\label{fig:CMG_v3y}
\end{figure}

The velocity field of \CMG\, as illustrated in Fig.~\ref{fig:CMG_v3y} here, is the basis of detailed analysis of the shape of its response current. We consider the current contribution of the green and yellow nodal structures in the top panel of Fig.5(a) of the main text. In Fig.5(b)(A) we mark several points for the drift $p_0$, and study the resulting drifted Fermi surface, as well as the change of the velocity occupied by it. The sharp increase from (i) to (ii) is a consequence of the rapidly varying velocity field, as soon as the Fermi surface is driven even slightly away from equilibrium. This sharp increase soon saturates and begins to fall as the flower petal is shifted away from the intense red patches situated near the touching points of the inner and outer chains [(ii) to (iv)]. Going from (iv) to (v), the current eventually changes sign because the intense blue patches are now being sampled, The current slowly becomes positive again [(v) to (vi)] when the flower petal is shifted to the outer chain [Fig.~\ref{fig:CMG}(b)].

In contrast to the current due to the green nodal structure [Fig.5(b)(A)], the gradient of the current is not as steep for the yellow structure [Fig.5(c)(A)]. Furthermore, the yellow circular ring also contribute exclusively negative velocity [Fig.5(c)(i-vi)] within the drift considered here, unlike the green one where the regions occupied by the ring vary in sign [Fig.5(b)(i-vi)]. These observations, together with the evolution of the velocity field similar to that discussed in the previous paragraph, can account for the current pattern of the yellow nodal structure.

\begin{figure}[H]
		\includegraphics[width=.9\linewidth]{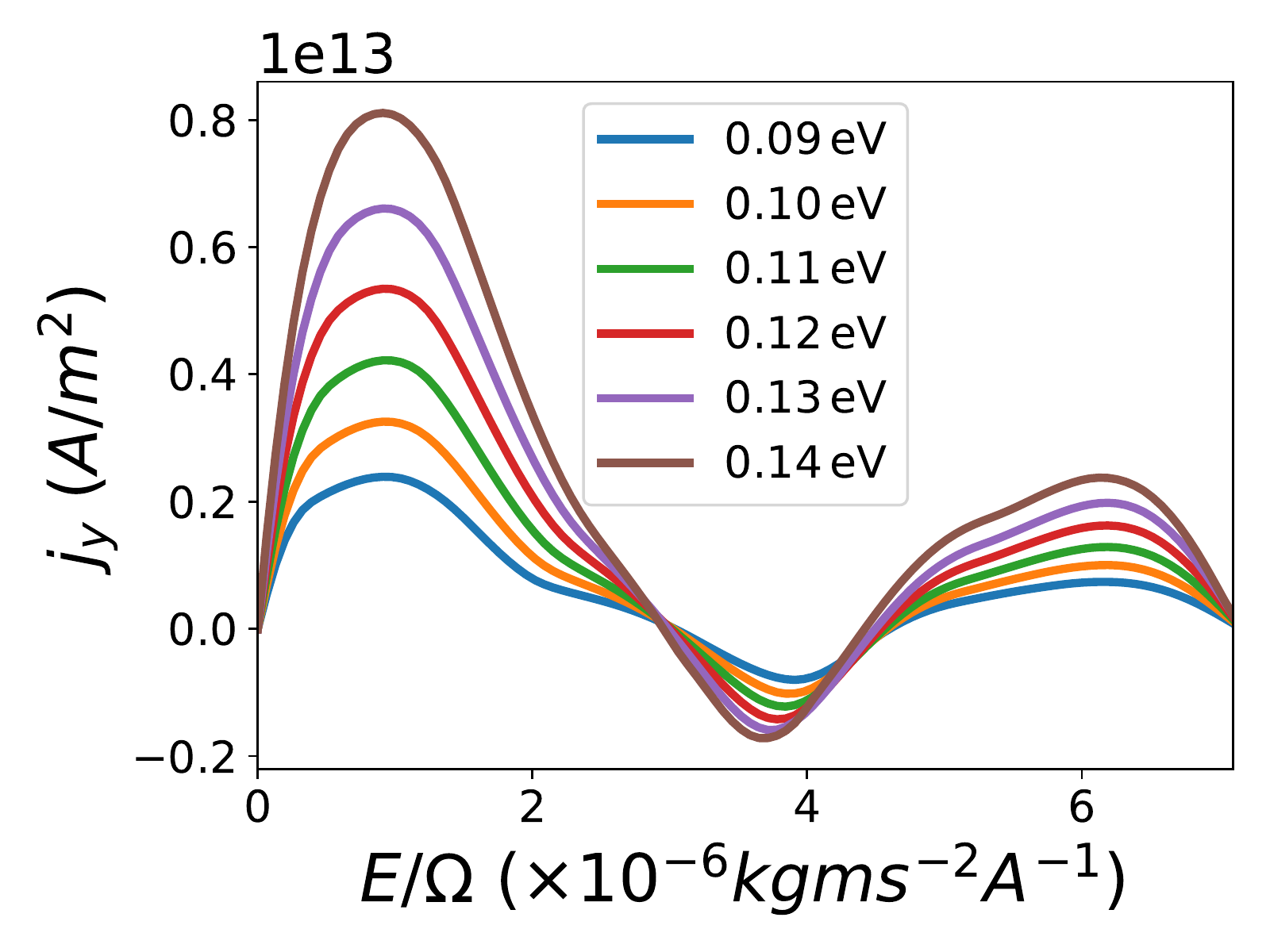}
		\caption{Response current of Co$_2$MnGa for different values of chemical potential $\mu$. A qualitatively similar curve persists, even though larger chemical potentials (thicker nodal regions) give rise to larger currents in general. }
		\label{fig:mu}
	\end{figure}

The above analysis demonstrates how the strong inhomogeneity of the velocity field, which results from the knotted links and touching loops, gives rise to its characteristic response. Indeed, on the two high symmetry planes displayed in Fig.~\ref{fig:CMG_v3y}(d-e), the flower petals are clearly discernible as contours of sign-changing points. As shown in Fig.~\ref{fig:CMG_v3y}(b), the Fermi surface of \CMG\, fits matchingly the void of the velocity field. By visualizing the velocity fields of simpler artificial nodal links (not shown), we observe that the more complicated a nodal network, the more often the velocity field switches direction. Our analysis holds best for small chemical potentials that results in thin NLs, even though the response curve usually remains qualitatively preserved as the chemical potential is increased (Fig.~\ref{fig:mu}), unless the NLs become so thick that they fuse and generate different nodal topologies. 

\subsection{Individual plots of the Fourier harmonics from HHG}
\begin{figure*}
	\subfloat[Ti$_3$Al]{
	\includegraphics[width=.33\linewidth]{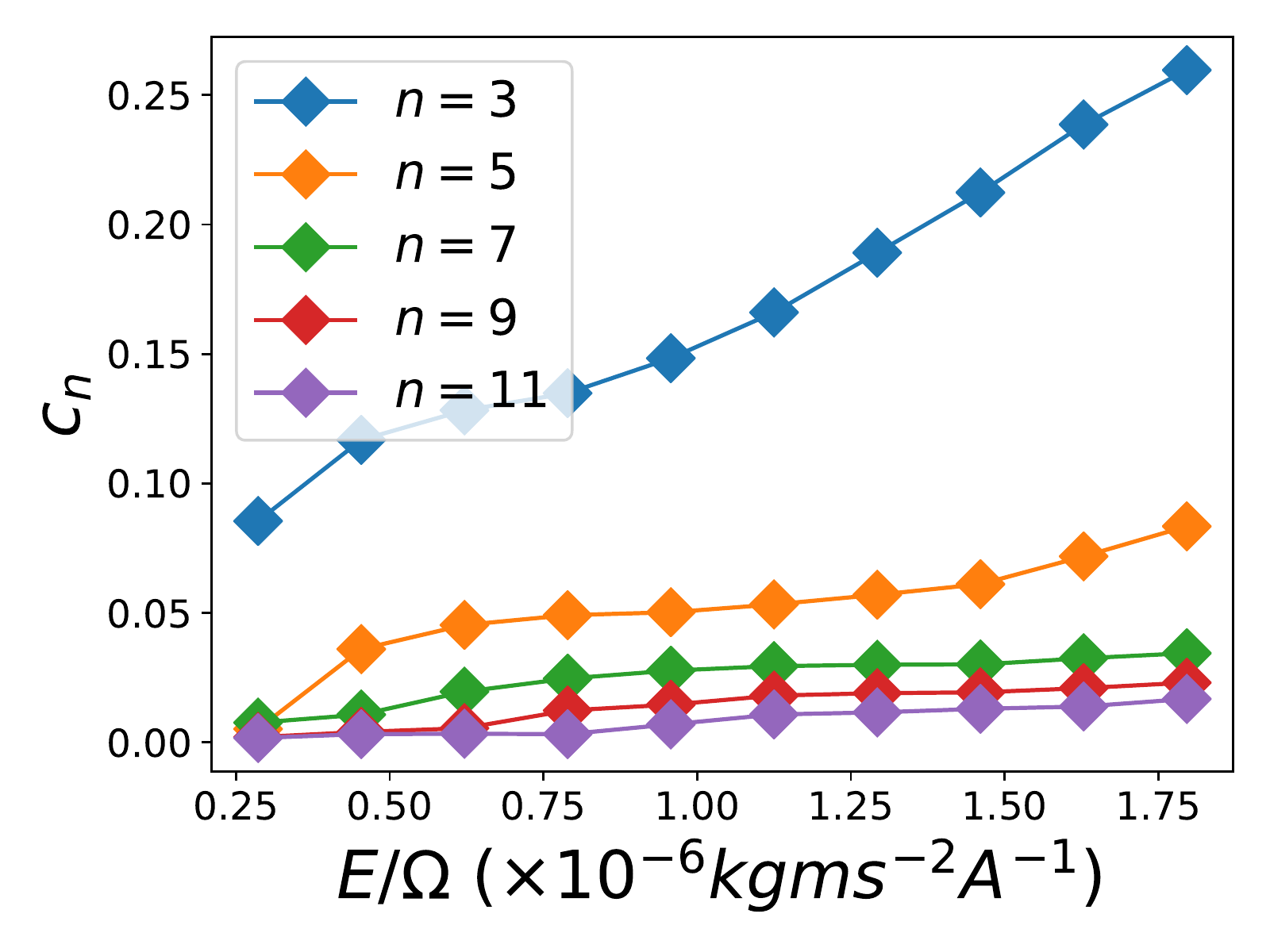}
	}
	\subfloat[YH$_3$]{
		\includegraphics[width=.33\linewidth]{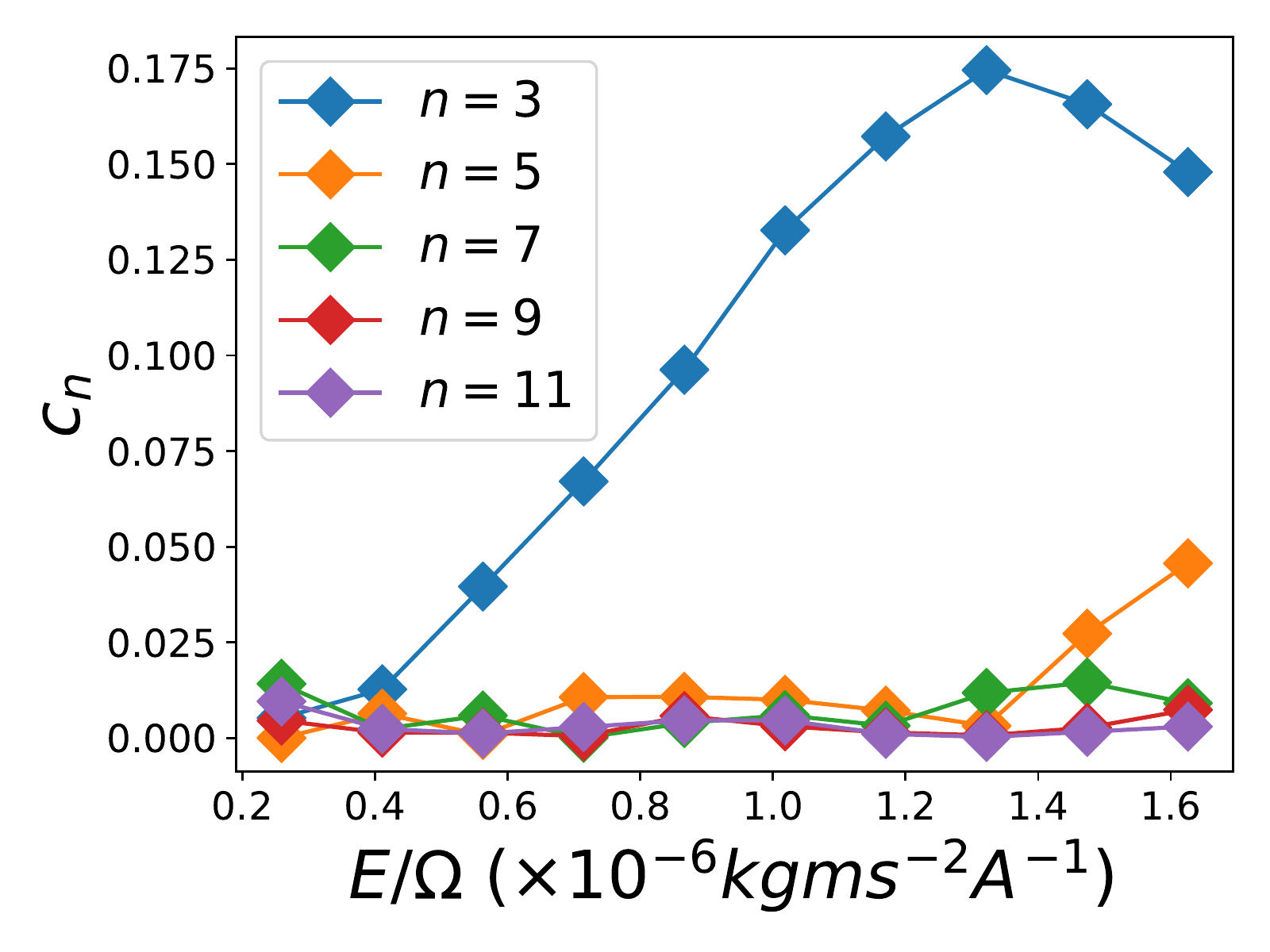}
	}
	\subfloat[Co$_2$MnGa]{
		\includegraphics[width=.33\linewidth]{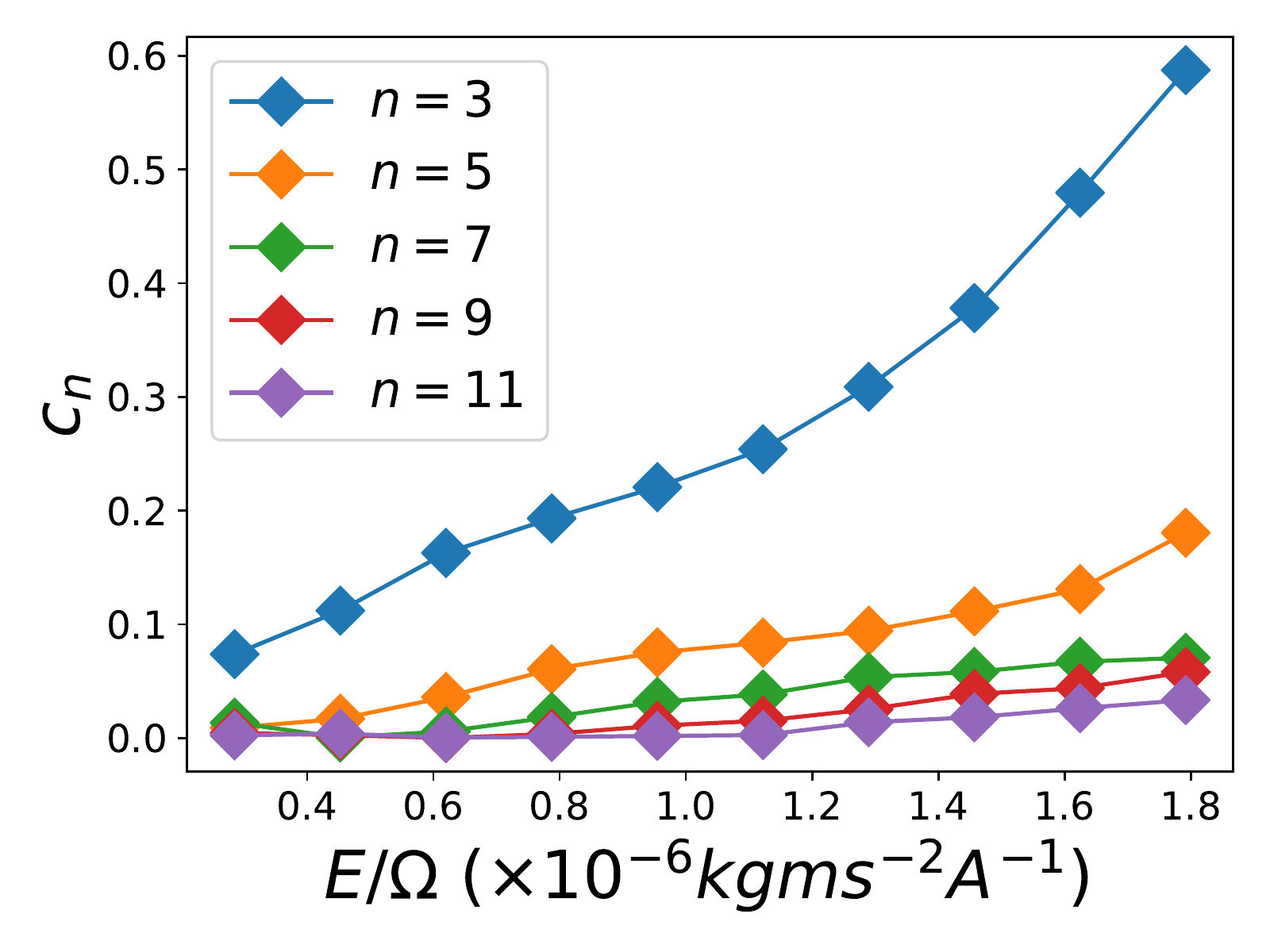}
	}
\caption{Higher harmonic generation (normalized by the first harmonic) for different field strength $p_0$ in (a) single nodal loop Ti$_3$Al, (b) triple inner chain YH$_3$, (c) nodal network Co$_2$MnGa. Note that the harmonic amplitudes $c_n$ for different materials are plotted in different scales across these figures; they are compared side-by-side in the same figure (Fig.~4e) in the main text.}
\label{fig:Fouriercomponents}
\end{figure*}

In Fig.~\ref{fig:Fouriercomponents}, we present more detailed data of the individual higher harmonics generated of our selected nodal materials in increasing level of complexity.

\subsection{Comments on our theoretical approach}
Here, we comment on our approach and connect it to a few experimental breakthroughs on HHG in solids. Throughout this work, we compute the electrical current by solving the semi-classical equations of motion (EOMs) to obtain the electronic velocity, which is then integrated over the phase space, weighted by the distribution function dictated by the Boltzmann equation.

From explaining the Wiedemann-Franz law to predicting the Bloch oscillation, such an approach has been hugely successful since the dawn of the quantum theory of solids~\cite{ziman2001electrons}. The semi-classical EOMs with ``anomalous velocity'' corrections are not only conceptually appealing but also predictively powerful in explaining diverse phenomena such as the various Hall effects and quantum oscillations~\cite{xiao2010berry}. As for the Boltzmann equation, it is expected to work best for weakly interacting systems which can be well described by the Landau theory of Fermi liquid~\cite{kubo1973boltzmann}. 
While the importance of correlation effects in nodal systems is certainly material-dependent, the fact that DFT+ARPES calculations of Co$_2$MnGa match so nicely the experimental results~\cite{belopolski2017three} may hint that a tight binding description of quasiparticles, and hence the application of Boltzmann equation, should be a good starting point. 

In general, solving the Boltzmann equation is highly nontrivial, complicated by the collision integral that assumes different forms and requires various approximation schemes depending on the scattering mechanisms. Confronted with this difficulty, progress can be made by working within the relaxation time approximation (RTA), where the collision integral is assumed to take the form $(F-f)/\tau$. This approximation is motivated by the consideration that the various types of scattering (electron-phonon, electron-impurity, electron-electron) will relax the system to a distribution $F$ (often chosen to be the Fermi-Dirac distribution) within a timescale $\tau$ which can be computed if a microscopic model for the scattering processes is given~\cite{ho2018theoretical}.

In our work, the collision integral is completely neglected. One way to justify this is as follows. Invoking the constant RTA, one has $(\partial_t + \dot{\bm{p}}(t)\cdot\partial_{\bm{p}})f=(F-f)/\tau$, where $f$ is the distribution function, $\bm{p}$ the momentum. In presence of a periodic electric field of frequency $\Omega$ [which enters the equation via $\dot{\bm{p}}(t)$], it is reasonable to assume a periodic steady state, provided the driving force is not too weak and not too slow compared to the dissipative relaxing term. Then, a periodic average of the equation implies that the LHS is of the order $O(\Omega\tau)$, and the RHS of order $O(1)$, so that for $\Omega\tau \gg 1$, the collision term may be ignored altogether. This conforms to the intuition that scattering is not so relevant in presence of a fast driving field~\cite{liu2017high}.

As with most physical sciences, theoretical descriptions of HHG in solids admit a varying degree of sophistication. For example, one may wish to examine the spatial distribution of the harmonics by taking a Wannier representation for the valence bands~\cite{osika2017wannier}, or to study the contributions of interband current by adding an interband polarization term~\cite{PhysRevB.91.064302}, or to account for the Coulomb interaction via a dephasing term~\cite{golde2006microscopic}. For the purpose of illustrating the HHG due to topologically enforced non-linearities in linked NLSMs, we considered only the intraband current within a non-interacting picture. This approach closely mirrors the theoretical model developed in the groundbreaking observation of HHG in bulk ZnO~\cite{ghimire2011observation}. There, a Peierl's substitution of a continuous wave (cw) electric field is performed on a single band of ZnO. The current is computed by multiplying the velocity with the electron charge and density. Simply by taking into account longer-range hoppings, qualitative features including the linear energy cutoff of the experimental HHG can already be captured. In our current expression, $J[\bold A(t)]=e\int_{\mathrm{BZ}} \frac{\mathrm{d}^3\bm{p}}{(2\pi\hbar)^3} f\left[\epsilon(\bm{p}-\bm{p}(t))	\right] \bm{v}(\bm{p})$, upon a coordination transformation, the electric field $\bm{p}(t)$ dependence turns up in the velocity, and the expression becomes similar to that in~\cite{ghimire2011observation}, with the exception that we integrate over the BZ instead of evaluating it at a certain momentum. We note that similar semi-classical approaches have subsequently been employed in~\cite{schubert2014sub} to discuss Bloch oscillations in HHG, as well as in~\cite{liu2017high} to study the role of Berry curvature in HHG of monolayer MoS$_2$. With regards to scattering and many-body effects, rather reassuringly, they should play only minor roles in HHG experiments~\cite{liu2017high,ndabashimiye2016solid}, in accordance with the intuition that most scattering cannot keep up with the fast driving field. Together with the fact that interactions are mostly insignificant in symmetry-protected topological materials, our theoretical approach may prove sufficient to describe HHG in nodal-line semimetals.

\end{document}